\documentclass[acmsmall,screen]{acmart}

\AtBeginDocument{%
  
    \setlength\abovedisplayskip{0pt}
    \setlength\belowdisplayskip{0pt}
    }
\usepackage{algorithm2e}
\usepackage{graphicx}
\usepackage{textcomp}
\usepackage{comment}
\usepackage{balance}
\usepackage[skip=3pt]{caption}

\usepackage{packages}

\setcopyright{acmlicensed}

\acmJournal{TOSEM} % ACM Transactions on Software Engineering and Methodology
\acmVolume{X}
\acmNumber{X}
\acmArticle{X}
\acmYear{2026}
\acmMonth{8}
\acmDOI{10.1145/XXXXXXX.XXXXXXX} % Replace with actual DOI
\acmISBN{978-1-4503-XXXX-X/2018/06}

\begin{document}

%%
%% The "title" command has an optional parameter,
%% allowing the author to define a "short title" to be used in page headers.
\title{Do They Do as They Are Told? Investigating Coding Agents' Adherence to Explicit Instructions in Context Files}

%%
%% The "author" command and its associated commands are used to define
%% the authors and their affiliations.
%% Of note is the shared affiliation of the first two authors, and the
%% "authornote" and "authornotemark" commands
%% used to denote shared contribution to the research.

\author{Gianmario Voria}
\email{gvoria@unisa.it}
\orcid{0009-0002-5394-8148}
\affiliation{%
  \institution{University of Salerno}
  \city{Fisciano}
  \country{Italy}
}

\author{Yutaro Kashiwa}
\email{yutaro.kashiwa@naist.jp}
\orcid{---}
\affiliation{%
  \institution{Nara Institute of Science and Technology}
  \city{Nara}
  \country{Japan}
}

\author{Gemma Catolino}
\email{gcatolino@unisa.it}
\orcid{0000-0002-4689-3401}
\affiliation{%
  \institution{University of Salerno}
  \city{Fisciano}
  \country{Italy}
}

\author{Fabio Palomba}
\email{fpalomba@unisa.it}
\orcid{0000-0001-9337-5116}
\affiliation{%
  \institution{University of Salerno}
  \city{Fisciano}
  \country{Italy}
}

%%
%% By default, the full list of authors will be used in the page
%% headers. Often, this list is too long, and will overlap
%% other information printed in the page headers. This command allows
%% the author to define a more concise list
%% of authors' names for this purpose.
\renewcommand{\shortauthors}{Voria et al.}

%%
%% The abstract is a short summary of the work to be presented in the
%% article.
\begin{abstract}
Developers increasingly rely on \textit{agent context files (ACFs)}, persistent, project-scoped artifacts such as \texttt{AGENTS.md}, to steer the behavior of autonomous coding agents without renegotiating expectations at every interaction. Yet this governance mechanism rests on an untested assumption: that instructions written into an ACF are actually carried out, rather than ignored or loosely followed. Existing studies describe what ACFs contain, but not whether individual instructions are followed, leaving open which guidance can be trusted. We address this gap through an empirical study of real-world ACF snapshots and their linked agent-generated commits: characterizing the explicit, actionable instructions ACFs encode, operationalizing them into executable measurement scripts, and using these scripts to assess instruction-level compliance. We find that most observable instructions can be operationalized and that agents follow ACF guidance more often than not, but adherence splits sharply between instructions followed consistently and those never respected, with differences across instruction categories and coding agents. This split shows where ACF guidance works and where it does not, pointing to instruction types that may need rewriting and to agents that may need enforcement beyond the ACF. These findings lay the groundwork for systematic approaches to specifying, verifying, and enforcing project-specific instructions in autonomous software development.
\end{abstract}

%%
%% The code below is generated by the tool at http://dl.acm.org/ccs.cfm.
%% Please copy and paste the code instead of the example below.
%%

%\ccsdesc[100]{Do Not Use This Code~Generate the Correct Terms for Your Paper}

%%
%% Keywords. The author(s) should pick words that accurately describe
%% the work being presented. Separate the keywords with commas.
%% A "teaser" image appears between the author and affiliation
%% information and the body of the document, and typically spans the
%% page.
%\begin{teaserfigure}
%  \includegraphics[width=\textwidth]{sampleteaser}
%  \caption{Seattle Mariners at Spring Training, 2010.}
%  \Description{Enjoying the baseball game from the third-base
%  seats. Ichiro Suzuki preparing to bat.}
%  \label{fig:teaser}
%\end{teaserfigure}

\received{20 February 2007}
\received[revised]{12 March 2009}
\received[accepted]{5 June 2009}

%%
%% This command processes the author and affiliation and title
%% information and builds the first part of the formatted document.
\maketitle
\section{Introduction} 
The emergence of \textit{autonomous coding agents} is progressively transforming how software engineering (SE) activities are organized and carried out~\cite{hassan2025agentic}. When provided with appropriate instructions and contextual guidance, these agents can complete sophisticated development tasks with relatively limited human oversight~\cite{li2025aiteammates}. Consequently, contemporary software development is increasingly moving toward a collaborative human--agent setting, where developers primarily specify goals, requirements, and constraints, while autonomous agents are responsible for translating them into executable development actions~\cite{hou2024large}.

In this context, a new class of software artifact has emerged: \textbf{agent context files (ACFs)}. These artifacts (e.g., \texttt{AGENTS.md}, \texttt{CLAUDE.md}, or \texttt{copilot-instructions.md}) act as agent-oriented counterparts of traditional README files, explicitly defining how an AI agent should behave within a given software project~\cite{chatlatanagulchai2025use, chatlatanagulchai2025agent}. ACFs encode project-specific knowledge, including architectural overviews, build and test procedures, coding conventions, and operational rules governing tool usage and external service interaction~\cite{chatlatanagulchai2025use}. Modern AI-assisted development tools, including Claude Code, OpenAI Codex, and GitHub Copilot, are designed to automatically load these files at the beginning of a session and use them as a stable reference when planning and executing code changes~\cite{chatlatanagulchai2025agent}. Consequently, ACFs play a central role in shaping how agents interpret a codebase, adhere to development practices, and perform non-trivial software engineering tasks.

\revised{Despite the growing adoption of ACFs, little empirical evidence exists on whether the instructions they encode are actually reflected in practice. The presence of guidance in a context file does not guarantee that an agent will follow it, and recent studies of agent-generated tests and of testing guidance in ACFs both name this question as future work~\cite{hora2026coding, ardic2026uncovering}. The question matters because ACFs are expected to act as persistent, project-specific sources of guidance~\cite{mohsenimofidi2025context} that let developers steer agent behavior in a controlled and predictable way. This expectation breaks down if instructions are systematically ignored, partially applied, or inconsistently interpreted, and recent evidence on agents' compliance with documented repository rules suggests that such failures do occur~\cite{yang2026firstlook}.}

\revised{Answering it also matters beyond confirming that ACFs work, because an instruction-level account of compliance has implications for two related problems in agent governance. Knowing which instructions are consistently followed and which are essentially never respected shows how effective specific forms of guidance are, and points to instruction types that may need reformulation or reinforcement beyond the context file. Differences in compliance across coding agents show how agents differ in interpreting project-level constraints, and expose instruction classes for which particular agents are less reliable. This gives a basis for agent-specific guidance and for verification mechanisms that complement natural-language instructions. An empirical account of instruction-level compliance thus yields design and engineering implications, not only a check on whether ACFs work.}

\revised{A fundamental challenge in addressing this problem lies in the nature of the instructions themselves. Existing studies describe ACF contents through high-level categories (e.g., testing, implementation, or quality) that aggregate heterogeneous forms of guidance with different meanings and implications~\cite{chatlatanagulchai2025agent}: a testing-related instruction, for instance, may mean running an existing test suite, adopting a specific framework, or generating new test cases. Such variation makes it difficult to reason about what it means for an agent to ``follow'' a given instruction type, or to identify observable evidence of such behavior in software artifacts. Hence, advancing our understanding of ACFs requires moving beyond high-level categorizations \textbf{toward a more precise characterization of the instructions they encode}---focusing on guidance that is explicit and actionable, directly stated in the ACF and prescribing concrete development behavior, and establishing whether each such instruction can be operationalized through measurable software artifacts.}

We address this gap by examining the relationship between explicit, actionable instructions in ACFs and the observable behavior of autonomous coding agents in real-world development. Rather than assuming the presence of guidance implies compliance\revised{, or treating compliance as a single aggregate figure,} we structure the investigation as a three-phase empirical pipeline: characterizing the instructions ACFs encode, establishing whether they are concretely measurable through software artifacts, and assessing how closely agents follow them, \revised{at a granularity fine enough to attribute compliance variation to specific instruction types and agents rather than to ACFs overall.}

\smallskip
\steResearchQuestionBox{\faBullseye \hspace{0.05cm} \textbf{Main Objective.} This study aims to empirically investigate whether instructions documented in agent context files are reflected in agent-generated software artifacts. Specifically, we aim to~(1)~characterize the explicit and actionable instructions encoded in real-world ACFs, (2)~establish whether such instructions can be operationalized through executable measurement scripts derived from their observable behavioral descriptors, and (3)~assess the extent to which these instructions are manifested in downstream changes produced by autonomous agents in real-world development settings.}
\medskip

We conduct a three-phase empirical study grounded in a dataset of 621 unique ACF snapshots linked to 6,541 agent-generated commits across 268 repositories. \revised{First, we combine manual calibration with LLM-assisted classification to extract and characterize explicit, actionable instructions from real-world ACFs, validating the results on a statistically representative sample. In the second phase, we generate executable measurement scripts via test-driven code generation: a code LLM independently samples candidate scripts and test fixtures from the instruction’s behavioral descriptor, and a script is accepted only if it passes a minimum number of fixtures and is corroborated by at least one independently generated script. In the third phase, we apply the validated scripts to the reconstructed commit dataset and analyze the resulting compliance observations through descriptive statistics and comparative tests across instruction categories and agents. Throughout, compliance denotes artifact-level evidence, produced by a script that operationalizes the instruction's own specified behavioral descriptor, that an applicable commit satisfies what the instruction prescribes.}

\smallskip
Applying this pipeline to the corpus, we find that most explicit, actionable, and observable instructions can be operationalized into executable measurement scripts and that, where compliance can be measured, agents follow ACF guidance more often than not---though adherence is far from uniform, splitting sharply between instructions followed consistently and those essentially never respected, with the gap shaped by both instruction category and coding agent. \revised{We use this split throughout the paper as evidence in its own right, indicating where current ACF content already functions reliably, where it does not, and which instruction classes and agents most warrant reinforcement beyond natural-language guidance alone.}

\smallskip
To sum up, this paper offers the following main contributions:
\begin{enumerate}
    \item A three-criteria characterization of the explicit, actionable, and observable instructions encoded in real-world ACFs, distinguishing genuinely verifiable guidance from high-level content categories used in prior work.
\smallskip
    \item An empirical evaluation of whether such instructions can be operationalized into executable measurement scripts and whether they are actually reflected in agent-generated commits, showing that compliance, while common, is neither guaranteed nor uniform across instruction types and agents.
\smallskip
    \item \revised{A discussion of the implications of these instruction-level compliance patterns for ACF authoring practices and for the engineering of agentic systems, identifying instruction types that may warrant reformulation and the classes of guidance for which verification beyond natural language appears most warranted.}
\smallskip
    \item A publicly available replication package\cite{replication_package}, including the extraction and classification pipeline, the generated measurement scripts, and the compliance dataset, to support future work on verifying and enforcing agent adherence to project-specific guidance.
\end{enumerate}
\smallskip

\section{Background and Related Work}\label{sec:relatedwork}

Software engineering practice is shifting from single-turn code completion toward autonomous coding agents that decompose high-level goals into multi-step plans and execute them with limited human intervention~\cite{hassan2025agentic}. Wang et al.~\cite{wang2025agents} characterize such systems through a perception--memory--action framework, in which agent behavior is governed in part by an explicit form of procedural memory encoded in the agent's configuration. This shift has broadened the range of activities agents are expected to perform, including automated program repair~\cite{bouzenia2025repairagent}, autonomous testing~\cite{feldt2023towards}, and agentic refactoring~\cite{horikawa2025agentic}, moving agents progressively closer to end-to-end ownership of development tasks.

Greater autonomy, however, introduces governance challenges. Empirical studies report integration defects and security risks specific to agentic settings~\cite{shao2024llms}, and consistently identify the absence of authoritative, project-specific knowledge as a primary failure mode~\cite{wang2025agents}. This failure mode is not merely theoretical: agents contributing to real-world pull requests still require substantial human revision because they fail to follow project-specific conventions~\cite{watanabe2025use}, indicating that capability alone does not guarantee conformance to project-level expectations. Maintaining human governance over autonomous agents has consequently emerged as a practical concern for professional developers. Huang et al.~\cite{huang2025professional} find that experienced developers deliberately resist delegating full control to agents, relying instead on explicit strategies---detailed prompts, staged plans, and continuous supervision---to retain authority over design and implementation decisions. Their findings show that governance is not a theoretical abstraction: practitioners actively constrain agent autonomy because unconstrained agents produce outputs that violate quality, safety, and design requirements. This motivates the need for mechanisms through which developers can encode such constraints persistently and systematically, rather than re-asserting them at every interaction.

\paragraph{Agent Context Files as Governance Artifacts}

In response to this need, developers have adopted persistent, project-scoped artifacts to communicate intent to agents~\cite{le2025impacts, shrivastava2023repository}. Agent context files (ACFs)---named \texttt{CLAUDE.md}, \texttt{AGENTS.md}, \texttt{.cursor/rules}, or \texttt{copilot-instructions.md} depending on the tool---capture guidelines, constraints, and conventions that remain active across sessions~\cite{chatlatanagulchai2025agent, chatlatanagulchai2025use}. Modern agentic tools load these files automatically at the start of a session and treat them as a stable reference throughout planning and execution~\cite{chatlatanagulchai2025agent}. Framed in terms of the perception--memory--action view introduced above, ACFs represent an explicit, developer-authored form of procedural memory: a mechanism through which project-specific knowledge is injected into the agent's context rather than inferred from the codebase alone.

Chatlatanagulchai et al.~\cite{chatlatanagulchai2025agent, chatlatanagulchai2025use} conducted the first large-scale empirical study of ACFs, analyzing 2,300 files across 1,925 repositories. They find that ACF content skews heavily toward functional concerns---build and run commands (62.3\%), implementation details (69.9\%), and architecture (67.7\%)---while non-functional requirements such as security (14.5\%) and performance (14.5\%) are rarely specified. Mohsenimofidi et al.~\cite{mohsenimofidi2025context} report substantial variation in how developers express instructions across projects, finding no established structural standard across the 466 projects examined. Galster et al.~\cite{galster2026configuringagenticaicoding} identify eight distinct configuration mechanisms ranging from static context files to executable subagents. \revised{Ardic et al.~\cite{ardic2026uncovering} study testing guidance in the ACFs of popular repositories with agent-authored pull requests. In a sample of 300 repositories, 249 include test commands, while guidance on test strategy, mocking, coverage, or naming is much less common.} Beyond characterizing content, recent work has begun to quantify the operational impact of ACFs: Lulla et al.~\cite{lulla2026impactagentsmdfilesefficiency} show that the presence of an \texttt{AGENTS.md} file is associated with a 28.6\% reduction in agent runtime and a 16.6\% reduction in token consumption.

Together, these studies establish ACFs as a widespread and influential class of software artifacts and confirm that their content measurably shapes agent behavior. Existing characterizations, however, remain descriptive: they document \textit{what} ACFs contain, typically through high-level content categories, rather than the extent to which the specific instructions they encode are actually followed.

\paragraph{Instruction Following and the Compliance Gap}

Whether the presence of an instruction in an ACF translates into compliant agent behavior remains largely unexamined. Adjacent evidence suggests it should not be assumed. Studies of AI-generated code quality show that functional correctness does not predict adherence to convention: code that passes all tests can still carry latent code smells and violate project-specific expectations~\cite{sabra2025assessing}, and even where agentic refactoring yields measurable structural improvements~\cite{horikawa2025agentic}, these gains are evaluated against generic quality metrics rather than against the developer-authored guidance the agent was given.

\revised{Instruction following has also been studied as a general LLM capability, including for code. \textsc{CodeIF-Bench}~\cite{wang2025codeifbench} measures whether generated code satisfies constraints stated in a single prompt, and \textsc{AgentIF}~\cite{qi2025agentif} extends this to long, multi-constraint system prompts in agentic settings. This work shows that instruction adherence is measurable and imperfect even for state-of-the-art LLMs. It evaluates instructions given within a single interaction, however, not a persistent, project-scoped artifact, and it does not look at the artifacts produced in real development.}

\revised{Studies of agents in real repositories come closer. Watanabe et al.~\cite{watanabe2025use} find that agent pull requests need substantial revision to meet project conventions, and Ehsani et al.~\cite{ehsani2026where} characterize at scale why agentic pull requests fail, without tying failures to the instructions an ACF documents. Hora and Robbes~\cite{hora2026coding} show that agents add mocks to tests more often than non-agents (36\% versus 26\% of test commits) and call for mocking guidance in agent configuration files, but they only check a few repositories anecdotally for whether such guidance is followed. Gloaguen et al.~\cite{gloaguen2026evaluating} evaluate whether ACFs help agents resolve issues, and find slightly lower task success for LLM-generated context files, slightly higher for developer-written ones, and about 20\% higher cost. Their trace analysis also shows that tools named in a context file are used far more often than tools that are not named. This measures tool usage in controlled issue-resolution tasks. We instead check, for individual instructions, whether their content shows up in the artifacts of real agent-generated commits.}

\revised{Closest to our setting, Yang et al.~\cite{yang2026firstlook} directly measure coding agents' compliance with documented rules. Their \textsc{RepoComplianceBench} covers repository-level AI-contribution policies (bans, disclosure requirements, verification gates) and uses a controlled set of 106 issues across four agents. It does not cover the broader development guidance that ACFs encode, nor compliance observed in real, mined commits.}

Existing work therefore establishes two main aspects. First, ACFs are widely adopted as behavioral governance artifacts. Second, agents do not reliably adhere to project conventions even without explicit, structured guidance. In this context, what remains missing is a systematic, artifact-grounded assessment of whether the specific instructions documented in an ACF are reflected in the commits an agent subsequently produces---and, more fundamentally, which of these instructions can be verified from produced artifacts at all.

\smallskip
\steSummaryBox{\faExclamationCircle \hspace{0.05cm} Novelty and Contribution.}{Prior work on ACFs is limited in two ways that this paper directly addresses. First, existing characterizations describe ACF content through high-level categories~\cite{chatlatanagulchai2025agent} that aggregate heterogeneous forms of guidance, leaving it unclear which instructions are explicit and actionable enough to support downstream verification. Second, no study measures whether individual ACF instructions are reflected in the commits agents produce in real projects. Our work addresses both gaps through a three-phase empirical pipeline: characterizing explicit and actionable ACF instructions, operationalizing them as executable measurement proxies, and assessing their manifestation in real agent-generated commits.}
\section{Empirical Study}\label{sec:design}
The \textit{goal} of this empirical study is to investigate the relationship between instructions encoded in agent context files and the observable behavior of autonomous coding agents in real-world software development, and to understand whether explicit guidance documented in these artifacts is reflected in the software modifications agents generate.
The \textit{purpose} is to characterize the types of instructions that can be empirically operationalized and to assess how far such instructions show up in downstream agent-generated commits.
The study is conducted from the \textit{perspective} of software engineering researchers and practitioners who want to understand how effective agent-oriented guidance is in practice. Rather than focusing on agent intent or internal reasoning, we take an artifact-centric view: we look at whether instructions are reflected in the observable software artifacts agents produce in real-world development.

\subsection{Research Questions}

Despite the growing use of agent context files to guide autonomous coding agents, it remains unclear what types of instructions they actually encode in a way that can be measured. Existing categorizations are typically too high-level and group together very different forms of guidance that cannot be directly measured in development artifacts~\cite{chatlatanagulchai2025agent}. To support a rigorous empirical investigation, we first need to identify which instructions are sufficiently \textit{explicit} and \textit{actionable}, meaning they are directly stated in the ACF and describe a concrete development behavior an agent can reasonably perform. Our \textbf{first RQ} builds a fine-grained dataset of explicit and actionable instructions from real-world ACFs, moving beyond high-level categories toward concrete guidance that supports downstream analysis.

\smallskip
\steResearchQuestionBox{\textbf{RQ\textsubscript{1}}. \emph{What types of explicit and actionable instructions are documented in agent context files?}}
\smallskip

Characterizing explicit and actionable instructions is necessary, but not enough for empirical measurement. We also need to establish whether each instruction can be verified through observable development artifacts, meaning whether compliance can be checked by inspecting produced or modified files, without inferring agent intent or internal reasoning. Treating observability as a purely classificatory label is risky: an instruction judged observable in the abstract may still turn out impossible to measure in practice. Our \textbf{second RQ} investigates whether observable measurement proxies can actually be built for the instructions identified in RQ\textsubscript{1}. For each instruction, we generate a measurable indicator that can determine compliance from commit-level artifacts. Instructions for which no such indicator can be built are considered non-operationalizable and excluded from downstream analysis. This process produces a refined dataset of instructions paired with validated measurement proxies, which is the empirical foundation for RQ\textsubscript{3}.

\smallskip
\steResearchQuestionBox{\textbf{RQ\textsubscript{2}}. \emph{To what extent can observable measurement proxies be operationalized for explicit and actionable instructions documented in agent context files?}}
\smallskip

Given a validated dataset of instructions with matching measurement proxies, the central question is whether autonomous coding agents actually follow the guidance in their ACFs. It remains unclear whether these instructions show up in the software artifacts agents produce during real development, or whether they are systematically ignored, partially applied, or inconsistently interpreted. Our \textbf{third RQ} investigates how far observable instructions are reflected in downstream agent-generated commits. For each instruction-commit pair, we apply the matching measurement proxy to the code changes and check whether observable evidence of compliance is present.

\smallskip
\steResearchQuestionBox{\textbf{RQ\textsubscript{3}}. \emph{To what extent are observable instructions documented in agent context files reflected in downstream agent-generated commits?}}
\smallskip

\begin{figure*}[t]
    \centering\includegraphics[width=\textwidth]{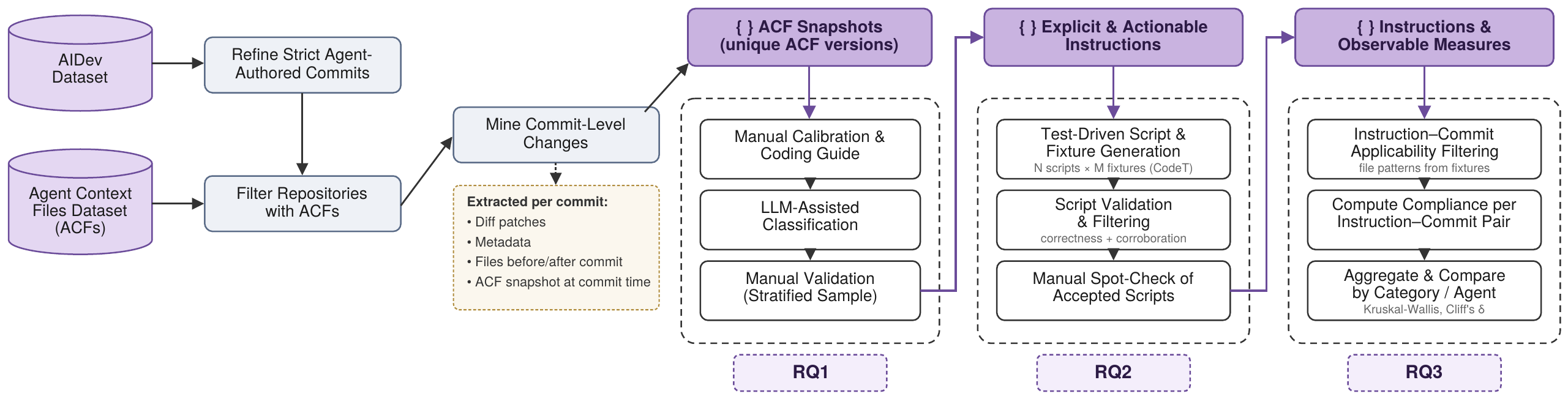}
    \caption{Overview of the research process.}
    \label{fig:study_design}
\end{figure*}

Figure~\ref{fig:study_design} summarizes the research method we followed. We first mined and combined data on agent context files and agent-generated code from two complementary datasets~\cite{chatlatanagulchai2025agent, li2025aiteammates}. We then built context file snapshots, unique versions of an agent context file defined by content and linked to the commits generated under that version, and classified their instructions using a mix of manual calibration and LLM-assisted qualitative analysis (RQ\textsubscript{1}). Next, we built a measurable proxy for each identified instruction, keeping only those for which a working indicator could be generated (RQ\textsubscript{2}). Finally, we applied the resulting measurement scripts to the reconstructed commit dataset and analyzed how far agent-generated code reflects the instructions in the associated ACF (RQ\textsubscript{3}).

\subsection{Data Collection} \label{sec:data_collection}
To investigate the relationship between agent context files and agent-generated development artifacts, we built a dataset by combining two complementary sources: (i)~an existing dataset of ACFs mined from GitHub repositories~\cite{chatlatanagulchai2025agent}, and (ii)~a large-scale dataset of agent-authored commits~\cite{li2025aiteammates}. Combining them required extra preprocessing and reconstruction steps to support fine-grained instruction-artifact analysis.

\subsubsection{Agent Context Files and Context Snapshots}
We started from the ACF dataset introduced in prior work~\cite{chatlatanagulchai2025agent}, which collects files from GitHub repositories based on the official naming conventions of widely used coding agents (e.g., \texttt{CLAUDE.md}, \texttt{AGENTS.md}, and \texttt{copilot-instructions.md}). We kept only repositories associated with three agents, Copilot, Claude Code, and OpenAI Codex, using the classification procedures from the original dataset for consistency. Agent context files change over time as developers update agent guidance. To capture the instructions available at the time of each commit, we built \textit{context snapshots}: unique versions of an ACF, deduplicated by content, so identical file contents map to the same snapshot and any change in content produces a new one. Each snapshot was then linked to every commit generated while that version was active. This let us associate each commit with the exact instruction set available to the agent at generation time, giving \textbf{676 unique context snapshots} reused across \textbf{10,763 commits} and spanning \textbf{307 repositories}. Context files varied a lot in size, with a median of 638 words (110 lines) and a maximum exceeding 16K words.

\subsubsection{Agent-Authored Commits}
To capture agent behavior in real-world development, we relied on the AIDev dataset~\cite{li2025aiteammates}, which attributes pull requests and commits to autonomous coding agents using signals such as bot accounts, branch naming conventions, and co-authorship markers. These signals differ across agents, so we distinguished two confidence levels. \textit{Strict agent commits} are those for which authorship can be inferred with high confidence from strong indicators, bot identities or unambiguous authorship metadata, following the same rules as AIDev. \textit{Non-human-suspected commits} are those with no clear sign of human authorship but with less certain attribution.

This distinction was needed because agents expose different authorship patterns: some appear as direct commit authors via dedicated bot accounts, others (e.g., Claude Code) appear as co-authors in commit messages, and tools such as GitHub Copilot may rely on indirect signals such as branch naming. All analyses reported in this paper use only the \textit{strict agent subset}, favoring attribution precision over recall; non-human-suspected commits are not analyzed further.

\subsubsection{Dataset Integration and Reconstruction}
The two datasets were combined at the repository level, keeping only commits from repositories with at least one available ACF. Starting from 21,513 commits, 8,600 had context information, of which 6,580 also had file-level change data. After excluding 39 planning and merge commits (planning commits, 3,149 overall, mostly carry no file changes and had already been removed by this filter), the analysis-ready dataset is \textit{6,541} commits across \textit{268} repositories. For each selected commit, we rebuilt the full development context by cloning the corresponding Git repository and extracting commit metadata, diff patches, file-level change information, and the content of all modified files before and after the commit. The before version came from the parent commit; the after version is the current commit.

All extracted artifacts were stored in a structured local format with automated completeness checks for missing commits, incomplete snapshots, and failed retrievals. All data and scripts are available in our replication package~\cite{replication_package}.

\subsubsection{Final Dataset}
Table~\ref{tab:dataset_characterization} summarizes the final dataset. The corpus spans a wide range of development activity: 62.2\% of commits modify source code, 21.1\% involve test files, and 16.4\% affect documentation, with smaller shares covering configuration and CI artifacts. It covers 26 programming languages, including C\#, TypeScript, Python, and Rust. With a median of 1 modified file per commit and a mean of 3.95, most commits are small, focused changes, which makes instruction-level compliance assessment tractable at the commit level.

\begin{table}[t]
\caption{Characteristics of the analyzed commit dataset.}
\label{tab:dataset_characterization}
\centering
\footnotesize
\rowcolors{1}{gray!15}{white}
\begin{tabular}{lr}
\rowcolor{black}
{\color{white}Metric} & {\color{white}Value} \\
\hline
Repositories & 268 \\
Commits & 6,541 \\
Changed files & 25,822 \\
Median changed files per commit & 1 \\
Mean changed files per commit & 3.95 \\
Maximum changed files in a commit & 1,056 \\
\hline
\hline
Commits touching source files & 4,067 \\
Commits touching test files & 1,381 \\
Commits touching documentation files & 1,072 \\
Commits touching configuration files & 329 \\
Commits touching CI files & 245 \\
Distinct programming languages & 26 \\
\hline
\end{tabular}
\end{table}

\subsection{\textbf{RQ\textsubscript{1}} --- Data Analysis}
\label{sec:rq1} To address \textbf{RQ\textsubscript{1}}, we used a three-phase approach combining manual qualitative analysis with LLM-assisted classification. The goal was to identify explicit and actionable instructions in agent context files, grounded in the high-level ACF categorization from prior work~\cite{chatlatanagulchai2025agent}. We retain that work's 19 categories as defined, including partially overlapping ones (e.g., \textit{Build} and \textit{Build and Run}, \textit{Management} and \textit{Project Management}), so that our results remain directly comparable to it rather than introducing a new taxonomy. The unit of analysis was the \textit{context snapshot}: a distinct version of an agent context file, deduplicated by content and linked to the commits generated while that version was active. After removing empty and malformed snapshots, the final population was \textbf{621 unique context snapshots}.

\subsubsection{Phase 1 --- Manual Calibration}
We started with an iterative manual analysis of randomly selected context snapshots to set stable extraction procedures. Rather than fixing a sample size up front, we proceeded incrementally following qualitative saturation criteria \cite{treude2024qualitative}: we continued until the extraction procedure, coding guidelines, and decision rules were stable enough to support prompt construction for automated classification. This phase covered \textbf{15 ACF snapshots}, after which no substantially new decision patterns appeared, so we considered the procedure stable. During this phase, we extracted and labeled candidate instructions, grounding them in the existing high-level ACF categories from prior work rather than building a new taxonomy. Each candidate instruction was judged against three criteria.
\begin{itemize}
    \item \textit{Explicitness}: the instruction must be directly and clearly stated in the ACF, not inferred from examples or context.
    \item \textit{Actionability}: the instruction must describe a concrete development behavior an agent can reasonably perform during software modification.
    \item \textit{Observability}: compliance must be checkable through produced or modified development artifacts, such as source code, documentation, configuration files, or repository structure, without inferring agent intent or internal reasoning.
\end{itemize}
The analysis focused only on behavioral guidance directed at agents; repository descriptions, architectural overviews, code examples, and other informational content were excluded. Disagreements and ambiguous cases were discussed among all authors and used to refine the coding guidelines.

\subsubsection{Phase 2 --- LLM-Assisted Classification}
Once the coding procedure was stable, we used an LLM annotator to classify instructions across the full dataset~\cite{tan2024large, ahmed2025can}. Manually coding all 621 snapshots would have taken substantial effort and introduced coder fatigue and inconsistency, so the LLM acted as a structured coding assistant rather than a source of ground truth.

\smallskip
\textbf{Annotation Pipeline.} Each context snapshot was processed independently using a fixed few-shot prompt with three parts: (i)~a system prompt defining the annotation task, the three coding criteria, and the output schema; (ii)~a curated set of few-shot examples from the calibration phase, covering clear inclusion cases, clear exclusion cases, and borderline cases; and (iii)~the target ACF content. Snapshots that exceeded the model's context budget were split into paragraph-level chunks, annotated independently, and merged back at the snapshot level. For each extracted instruction, the model produced a JSON object recording the raw instruction text, high-level category, normalized label, coding decisions for each criterion, a \textit{candidate observable behavior} field describing how compliance could be verified in produced artifacts, an inclusion decision, and a mandatory justification note.

\smallskip \textbf{Prompt Engineering.}
The prompting strategy was developed iteratively through pilot runs on the calibration examples. Early versions systematically over-included vague or process-oriented statements as observable. The final version strengthened the observability criterion: an instruction now counts as observable only if a reasonably simple verification function could inspect the generated or modified files and determine compliance, with no access to agent reasoning, command execution, or internal workflow state. All prompt versions, model configurations, execution scripts, and few-shot examples are available in the replication package~\cite{replication_package}.

\smallskip \textbf{Model Selection.} We evaluated two candidate annotators, \textit{Ministral 3 14B Reasoning} and \textit{Qwen 3.5 9B}, both run locally via LMStudio with fixed inference parameters (32k-token context, temperature 0.1, top-\textit{p} 0.85, GGUF \texttt{Q4\_K\_M} quantization). A cross-model comparison on a shared set of annotations gave an overall agreement rate of 74.8\%, with 79.5\% include-agreement and 76.1\% observable-label agreement. Manual inspection of the disagreements showed that Ministral systematically over-extracted candidate instructions, accepting vague or process-oriented statements that Qwen correctly excluded. We therefore chose \textit{Qwen 3.5 9B} as the primary annotator for the final extraction.

\subsubsection{Phase 3 --- Manual Validation}
We did not assume LLM-generated annotations were equivalent to human judgment. Following established practice for LLM-based annotation~\cite{tan2024large, ahmed2025can}, we validated the automated classifications against the predefined coding criteria using a human annotator on a statistically representative sample.

\smallskip \textbf{Sample Selection.} The validation population was the 621 processed snapshots. Standard finite-population sampling at a 95\% confidence level and 10\% margin of error gave a validation sample of \textbf{84 context snapshots} (13.5\% of the population). We set the sampling unit at the snapshot level rather than the instruction level on purpose: validating a whole snapshot, rather than isolated instructions sampled across snapshots, lets the annotator check whether the extraction procedure correctly identifies all relevant instructions in a given ACF, including omissions that an instruction-level sample would miss. To ensure representativeness, we used proportional stratified random sampling based on (i)~the agent associated with the snapshot (Copilot, Claude Code, OpenAI Codex) and (ii)~instruction density (low: 1--10 instructions; medium: 11--30; high: $>$30).

The 84 sampled snapshots together contain \textbf{2,202 individual instructions} for validation, both included and excluded by the LLM, corresponding to 14.8\% of all 14,902 extracted instructions. This is the raw count of candidates the LLM extracted from these snapshots, before human review. The results (Section~\ref{sec:rq1_validation}) also report a narrower count of 2,196 candidates that received a recorded validation decision, and a still-narrower count of 2,114 instructions confirmed as genuine. Each of these three counts excludes a different subset (undecided candidates, and false positives, respectively) and each is used where it fits that specific analysis. This instruction-level coverage is well above the minimum sample size needed for 95\% confidence and 10\% margin of error at the instruction level for a population this size (96 instructions), giving strong grounding for the instruction-level metrics we report.

\smallskip \textbf{Validation Protocol.}
The validation was done by the first author, who had developed and applied the coding guide during calibration. Following \citet{ortloff2023different}, we used a single annotator for this well-bounded classification task: their study shows that for tasks with a narrow interpretive range, one experienced coder can be enough, and may even give higher internal validity than multi-coder setups with reconciliation noise. This fits our setting because each of the three coding criteria is a binary decision grounded in explicit textual evidence, and because the coding guide had already been refined with all coauthors before validation began. Low-confidence decisions during validation were still discussed with all coauthors until we reached a shared interpretation.

The resulting data were used to compute extraction-level precision, recall, and F1-score, and label-level accuracy for explicitness, actionability, observability, inclusion decisions, and category assignment. We also validated the \textit{candidate observable behavior} field: high accuracy here shows that the LLM-generated behavioral descriptors are reliable enough to use as input for the measurement proxy generation step in \textbf{RQ\textsubscript{2}}.

\subsubsection{RQ\textsubscript{1} Output} The output of \textbf{RQ\textsubscript{1}} is a validated dataset of explicit and actionable instructions extracted from agent context files, grounded in the existing high-level ACF categorization. For each instruction, the dataset records its textual formulation, category, per-criterion coding decisions, and candidate observable behavior. This dataset gives a fine-grained picture of the concrete guidance documented in ACFs and how it is spread across repositories, agents, and instruction categories. Instructions validated as observable, together with their candidate behavioral descriptors, are the foundation for \textbf{RQ\textsubscript{2}}.

\subsection{\textbf{RQ\textsubscript{2}} --- Data Analysis} \label{sec:rq2}

To address \textbf{RQ\textsubscript{2}}, we investigated whether the explicit and actionable instructions from \textbf{RQ\textsubscript{1}} can be turned into executable measurement scripts. Each instruction in our dataset is, by construction, explicit and describes a concrete development behavior verifiable through software artifacts. This makes each instruction suited to a functional specification: a well-defined procedure that takes commit-level artifacts as input and returns a binary compliance signal as output.

Turning a natural language specification into such a procedure is a code generation task, and LLMs have shown strong, well-documented capability at this~\cite{jiang2026survey}. We treat an instruction as operationalizable only if a correct measurement script can be generated and validated for it; instructions for which this fails are excluded from the final \textbf{RQ\textsubscript{3}} analysis.

\subsubsection{Measurement Script Specification}

Each measurement script is generated independently for a single instruction. We first considered grouping instructions that share a topic (e.g., naming conventions, indentation style) under a shared canonical representative before generation, on the idea that behaviorally equivalent instructions should map to the same measurement procedure. A closer look at the \textbf{RQ\textsubscript{1}} dataset showed that this equivalence rarely holds: instructions on the same high-level topic often prescribe different, mutually exclusive requirements. For example, one instruction may require camelCase for method identifiers, another PascalCase for type names, and a third snake\_case for file names, all under ``naming conventions.'' Collapsing these into one canonical representative would either drop the specific requirement each instruction mandates, or require per-instance parametrization of a shared script, which carries the same overhead as generating a script per instruction directly while adding a new point of failure (an instruction wrongly assigned to a template family, or a parameter wrongly extracted). We therefore generate one measurement script per instruction.

For each instruction retained by \textbf{RQ\textsubscript{1}} (explicit, actionable, and observable), we use its \textit{candidate observable behavior} field as the functional specification for script generation. This field, produced during \textbf{RQ\textsubscript{1}} extraction and confirmed on the manually validated sample, describes in concrete terms how compliance can be verified by inspecting files modified or added in a commit. For an instruction prescribing camelCase naming for method identifiers, for example, the candidate observable behavior states that all method names added or modified in the commit must match a camelCase pattern.

\subsubsection{Test-Driven Measurement Script Generation}

Script generation follows a test-driven approach, building on evidence that pairing a formal specification with independently generated test cases substantially improves the correctness of LLM-generated code~\cite{mathews2024test, fakhoury2024llm}. Rather than generating one script and repairing it against failures, we adopt the sampling-and-selection strategy from Chen et al.'s \textsc{CodeT}~\cite{chen2022codetcodegenerationgenerated}: for each instruction, the same code LLM independently generates $N$ candidate measurement scripts and $M$ candidate test fixtures, where each fixture is a pair of synthetic commits, one compliant and one non-compliant, synthesized from the candidate observable behavior specification. Each synthetic commit uses the same four-field \texttt{commit} object that scripts take as input, \texttt{files\_before}, \texttt{files\_after}, \texttt{message}, and \texttt{branch}, so a script is validated on exactly the input format it later receives for real commits. Fixtures carry an intended label since they are generated from an explicit compliant/non-compliant prompt, but we do not treat this label as ground truth without checking it, since the LLM can misread an ambiguous instruction when building a fixture, mirroring the test-case correctness issues reported for LLM-generated tests in general~\cite{chen2022codetcodegenerationgenerated}.

Every candidate script is run against every candidate fixture. In \textsc{CodeT}'s original dual execution agreement procedure~\cite{chen2022codetcodegenerationgenerated}, scripts that give exactly the same verdicts across all fixtures form a consensus group, and the group scoring highest on the product of its size and the number of fixtures it passes is selected. We retain this principle, that a script is trusted only when independently generated scripts agree with it on independently generated tests, but replace exact-verdict grouping with the overlap-based acceptance criterion described in Section~\ref{sec:rq2_acceptance}, because exact agreement proved too sparse at our sampling budget. $N$ and $M$ are small compared to the sampling budgets used in general-purpose code generation benchmarks, since the measurement scripts here target a narrower and more uniform class of checks than the algorithmic problems \textsc{CodeT} was originally evaluated on. We therefore use the exhaustive variant of the agreement procedure, comparing all script-fixture pairs directly rather than sampling from them.

%\topar{Model and configuration TBD. A code-specialized model (e.g., a recent code-instruct variant) is preferable. Report: model name, context window, temperature, quantization if local. Sampling budget ($N$ scripts, $M$ fixtures) TBD via pilot calibration on a stratified subsample; report the value(s) chosen and the stability of the top consensus group across a range of budgets.}
We use Qwen2.5-Coder-14B-Instruct as the generation model, matching the local execution setup used for instruction extraction in \textbf{RQ\textsubscript{1}} (LMStudio, 32k context window, GGUF Q4\_K\_M quantization) to keep the infrastructure consistent and fully reproducible on the same hardware. We confirmed this configuration ran successfully on the available hardware before the full \textbf{RQ\textsubscript{2}} run. Unlike the extraction stage, which favors deterministic, high-precision output (temperature 0.1), script and fixture generation needs diversity across independently sampled candidates for the dual execution agreement mechanism to work, so we set temperature to 0.8, the value Chen et al.~\cite{chen2022codetcodegenerationgenerated} validated for this purpose, and kept top-p at 0.85.

We fixed the sampling budget to $N = M = 10$: ten candidate scripts and ten candidate fixture pairs per instruction, compared exhaustively (100 script-fixture pairs). This is deliberately smaller than the budgets in \textsc{CodeT}'s original evaluation (50--1,000, depending on benchmark difficulty), which targeted open-ended competitive programming problems with many structurally different correct solutions~\cite{chen2022codetcodegenerationgenerated}. In our setting, each instruction comes with an explicit candidate observable behavior specification that narrows the space of correct implementations considerably, so a smaller candidate pool is enough for the corroboration mechanism to reach a stable consensus. We did not tune $N$ and $M$ separately (the pilot described in Section~\ref{sec:rq2_acceptance} concerned the acceptance criterion, not the budget); the operationalization rate reported in \textbf{RQ\textsubscript{2}} (the share of instructions for which a script is accepted) works as the direct empirical check on whether the fixed budget is adequate.

\subsubsection{Script Validation and Filtering} \label{sec:rq2_acceptance}

A script is accepted only if it clears two bars: a minimum-correctness bar (it must pass a minimum number of fixtures outright) and a corroboration bar (at least one other independently generated script must agree with it on a large share of the fixtures it passes). Neither bar alone is enough. The first alone would accept a script that happens to satisfy one fixture by chance, and the second alone, agreement with no floor on correctness, would accept two scripts that are consistently wrong the same way. Requiring both keeps \textsc{CodeT}'s core principle, that an accepted solution must be corroborated and not just self-consistent, while staying feasible at the sampling budget used here ($N = M = 10$). In a pilot run of the pipeline on a subset of instructions % TODO: add pilot size (n instructions) and the share accepted under exact-match agreement
before the full \textbf{RQ\textsubscript{2}} run, exact-match convergence across independently generated scripts and fixtures essentially never happened, leaving almost every instruction without an accepted script.

We set the minimum-correctness bar to two fixtures passed and the corroboration bar to 60\% overlap with the best-performing script's passed-fixture set. Both values were fixed on that same pilot rather than through a separate calibration study: a lower correctness floor risks accepting a script that only happens to satisfy a single fixture, while a much higher overlap threshold brings back the sparsity problem that led us to abandon exact-match agreement in the first place. Instructions for which no script clears both bars are excluded from the \textbf{RQ\textsubscript{3}} analysis. Rather than falling back to manually authored scripts, we treat exclusion as final, keeping one fully automated generation pipeline. Each exclusion is logged with the instruction's high-level category, originating ACF, and coding agent, so we can check exclusions for patterns rather than treat them as missing at random. This pass/fail rule is our operational definition of operationalizability: an instruction resists operationalization if no script clears both bars within the sampling budget, regardless of whether it was classified as observable in \textbf{RQ\textsubscript{1}}.

Relaxing exact-match agreement to overlap-based corroboration lowers the bar for what counts as agreement, which could in principle raise the risk of accepting a script that is subtly wrong. To check this, we manually inspect a sample of accepted scripts and verify that each one correctly implements its instruction's candidate observable behavior, using the same single-annotator validation protocol as \textbf{RQ\textsubscript{1}}. We report the agreement rate between this manual check and the automated acceptance decision as a direct, empirical measure of how reliable the accepted scripts are, alongside the operationalization rate itself.

We size the spot-check sample using the same finite-population-corrected Cochran formula, 95\% confidence level, and 10\% margin of error used for \textbf{RQ\textsubscript{1}}'s 84-snapshot validation sample (Section~\ref{sec:rq1}), applied here to the population of 4{,}123 accepted scripts instead of context snapshots. This gives a target sample size of 94. We stratify by category and allocate the 94-script target proportionally to each category's share of accepted scripts, using the largest-remainder method to hit the target exactly and a floor of 1 script per category so none is left entirely unchecked. Within each category, scripts are drawn uniformly at random without replacement, using a fixed random seed for reproducibility.

Each sampled script is reviewed by one author alongside its originating instruction's raw text and \textit{candidate observable behavior} specification, and judged \emph{agree} (the script correctly implements the specified behavior) or \emph{disagree} (it does not), with free-text notes for any judgment that is not immediately obvious from the two-way label alone. We refer to an accepted script together with the scripts that corroborate it as its consensus group.

\subsubsection{RQ\textsubscript{2} Output}

The output of \textbf{RQ\textsubscript{2}} is a validated set of instructions paired with executable measurement scripts. For each instruction, the dataset records the candidate observable behavior descriptor, the generated script, the fixture set used for validation, and the operationalizability verdict. This dataset is the measurement infrastructure used in \textbf{RQ\textsubscript{3}}.

Beyond the script artifacts themselves, we report the \textit{operationalization rate}, the share of \textbf{RQ\textsubscript{1}}'s explicit, actionable, and observable instructions for which a validated script was obtained, both overall and broken down by \texttt{high\_level\_category} and by coding agent. This breakdown shows which classes of instructions are measurable at all, which directly bounds what \textbf{RQ\textsubscript{3}} can say about adherence. If the operationalization rate were markedly lower for substantive categories (e.g., testing, security, architecture) than for surface-level categories (e.g., formatting, naming), that would mean \textbf{RQ\textsubscript{3}}'s adherence results, however precise, describe compliance for a subset of instructions skewed toward superficial conventions rather than the full space of explicit and actionable directives in ACFs. We report this breakdown as a required result of \textbf{RQ\textsubscript{2}}, independent of \textbf{RQ\textsubscript{3}}'s findings, and discuss what it means when interpreting the \textbf{RQ\textsubscript{3}} results.

\subsection{\textbf{RQ\textsubscript{3}} --- Data Analysis} \label{sec:rq3}

To address \textbf{RQ\textsubscript{3}}, we applied the measurement scripts from \textbf{RQ\textsubscript{2}} to the mined commit dataset and checked whether the instructions in each ACF are reflected in the matching agent-generated code changes. The analysis works at the level of individual (instruction, commit) pairs: for every commit generated while a given context snapshot was active, we consider each instruction from that snapshot for which \textbf{RQ\textsubscript{2}} produced an accepted measurement script. Instructions without an accepted script are excluded from this analysis, by \textbf{RQ\textsubscript{2}}'s own definition of operationalizability. This is the coverage boundary discussed in Section~\ref{sec:rq2_outcomes} and revisited in Section~\ref{sec:rq3-results}.

\subsubsection{Instruction--Commit Applicability}

\smallskip \textbf{Design.} Not every operationalized instruction is worth checking against every commit. A naming-convention instruction for Python method identifiers has nothing to say about a commit that only touches a YAML configuration file. Running its script against such a commit is not a neutral no-op: LLM-generated scripts tend to be permissive by default. They are written to look for evidence of a violation and report compliance when none is found, rather than requiring positive evidence that the check actually applied. So running a script against a commit it has nothing to check skews toward compliance with no real evidence behind it, rather than a neutral null result (Section~\ref{sec:rq2_diagnostics} reports how common this default is among accepted scripts). Evaluating every instruction against every commit without filtering would inflate every adherence figure in this study with this kind of empty compliance, not genuine adherence. An applicability filter, run before script execution, is a methodological necessity, not an optional extra.

We first considered setting applicability from hand-authored, category-level rules (e.g., ``naming-convention instructions apply to commits touching source files,'' ``testing instructions apply to commits touching source or test files''). We reject this approach: it needs inventing and validating a new mapping from category to file type at a level of detail too high-level to be reliable. A category such as \textit{Testing} covers instructions targeting Python test files, Go test files, JavaScript configuration files, and CI pipeline definitions alike, so no single category-level rule captures this.

\smallskip \textbf{Implementation.} We instead derive applicability directly from each instruction's own accepted evidence, rather than from a newly written rule. Every accepted instruction in \textbf{RQ\textsubscript{2}}'s output carries the fixture set its script was validated against (\texttt{fixtures\_used}), each fixture a realistic before/after commit example built specifically for that instruction. For each accepted instruction, we take the union of file paths in the \texttt{files\_before} and \texttt{files\_after} of all fixtures in its winning consensus group, and pull out from each path either its extension (e.g., \texttt{.py}, \texttt{.json}) or, for extensionless conventional filenames (e.g., \texttt{Dockerfile}, \texttt{Makefile}), its exact basename. This gives a small set of file patterns per instruction, grounded in the same artifacts \textbf{RQ\textsubscript{2}} already validated the script against, rather than a newly introduced heuristic. A commit is applicable to an instruction if it touches at least one file matching one of these patterns. Instructions whose fixtures have empty \texttt{files\_before} and \texttt{files\_after} throughout, meaning their compliance criterion concerns only the commit message or branch name, are treated as universally applicable, since every commit has a message and a branch regardless of which files it touches.

Instruction-commit pairs that fail the applicability check are excluded from evaluation entirely; they contribute no observation, compliant or otherwise. We report, per instruction, the number of applicable commits found, and separately flag any accepted instruction with zero applicable commits in the mined dataset. These cannot contribute an adherence observation despite being operationalizable, and are excluded from the statistical analysis in Section~\ref{sec:rq3-stats} on that basis. This is a further, separate loss beyond \textbf{RQ\textsubscript{2}}'s operationalization funnel, and we report it with the same transparency.

\subsubsection{Instruction--Commit Compliance Evaluation}
\revised{We use \textit{compliance} to mean the presence, in an applicable commit's artifacts, of the behavioral evidence a script derives directly from the instruction's own \textit{candidate observable behavior} (Section~\ref{sec:rq2}). That field is written from the instruction's own text during \textbf{RQ\textsubscript{1}} extraction, and the script is generated and validated against fixtures grounded in it before ever being applied to a real commit, so a compliance verdict reflects what the instruction itself specifies as satisfying behavior.}
For every instruction-commit pair that passes the applicability filter, we run the instruction's accepted script against that commit's artifacts and record a binary compliance signal. The script's input is exactly the four-field \texttt{commit} object defined in Section~\ref{sec:rq2}, \texttt{files\_before}, \texttt{files\_after}, \texttt{message}, \texttt{branch}, populated from the mined commit dataset rather than a synthetic fixture. Scripts do not additionally receive a line-level diff or patch; they were generated and validated only against full before/after file content, and extending the contract after the fact to include diff-level information would invalidate the already-generated and already-validated script population. This means instructions whose true compliance criterion depends on what was specifically added or removed, rather than the file's final state, are evaluated on a weaker signal than they ideally would be. We treat this as a limitation of the current design, discussed further under construct validity in Section~\ref{sec:threats}, rather than fix it by regenerating scripts.

\subsubsection{Aggregation and Statistical Analysis} \label{sec:rq3-stats}

We do not compare instruction categories or coding agents directly on the raw (instruction, commit) observations, because these observations are not independent: commits from the same context snapshot share the same instruction set, and instructions evaluated within the same commit share the same agent and repository. Treating each pair as an independent sample would understate variance and risk finding significance that is not really there. We instead aggregate to one compliance rate per instruction, the share of its applicable commits in which it was followed, before making any cross-group comparison, so each instruction contributes exactly one observation regardless of how many commits it was evaluated against.

We report descriptive statistics on this instruction-level compliance-rate dataset: the overall mean compliance rate, the rate by \texttt{high\_level\_category}, and the rate by coding agent (Copilot, Claude~Code, OpenAI~Codex), matching the reporting structure used for operationalization rate in \textbf{RQ\textsubscript{2}}. To test whether compliance rate differs across categories, we compare the distributions of per-instruction compliance rates using the Kruskal-Wallis test, a good fit given the expected non-normality of rate data bounded in $[0,1]$ and the unequal group sizes across categories. Where this test is significant, we run post-hoc pairwise Mann-Whitney U tests between categories with Bonferroni correction for the number of comparisons, and report Cliff's delta as the effect size for each significant pair, so the size of a difference is reported independently of the sample size behind it. We repeat the same procedure with coding agent as the grouping variable, to test whether adherence patterns differ by agent and not only by instruction category.

\revised{Many instructions are evaluated against very few applicable commits: the median evaluated instruction has only 3, and 540 (26.3\%) have exactly 1, which forces their individual compliance rate to 0\% or 100\% regardless of true adherence. To check whether this thin-evidence tail is driving the bimodal split and the two cross-group comparisons, we run a \textit{sensitivity analysis} repeating the full statistical procedure above restricted to instructions with at least 3, and separately at least 5, applicable commits, and report whether the comparisons found significant in the full sample stay significant as this low-evidence tail is excluded.}
\section{Analysis of the Results}
\label{sec:results}

\subsection{RQ\textsubscript{1} --- What types of explicit and actionable instructions are documented in agent context files?}
\label{sec:rq1-results}

We answer \textbf{RQ\textsubscript{1}} by applying the classification pipeline from Section~\ref{sec:rq1} to all 621 context snapshots. Section~\ref{sec:rq1_distribution} reports how the extracted instructions split across the three coding criteria: explicitness, actionability, and observability. Section~\ref{sec:rq1_validation} then reports the manual validation we use to check how reliable this automated classification is.

\subsubsection{Instruction Extraction Across the Three Coding Criteria}
\label{sec:rq1_distribution}

Table~\ref{tab:rq1_funnel} shows how the 14{,}902 candidate instructions extracted from the 621 context snapshots (24.0 per snapshot on average) are classified by the three coding criteria. It also reports the same breakdown for the 84-snapshot validated sample, restricted to the instructions confirmed as genuine during manual validation (Section~\ref{sec:rq1_validation}). Applied in sequence to the full corpus, the criteria narrow the candidates to 10{,}324 (69.3\%) that are explicit, 8{,}726 (58.6\% of all candidates) that are also actionable, and 6{,}552 (44.0\% of all candidates) that are also observable. This final set, which satisfies all three criteria, forms the \textbf{RQ\textsubscript{1}} dataset. It covers 592 of the 621 snapshots (95.3\%), and 615 snapshots (99.0\%) contain at least one explicit and actionable instruction. Explicitness is the first and, in absolute terms, the largest filter, discarding 30.7\% of all candidates. Among the criteria applied to already explicit instructions, observability is the more selective: it removes about a quarter of the explicit-and-actionable instructions at the corpus level (24.9\%) and close to half at the sample level (48.9\%), whereas actionability removes 15.5\% and 18.4\% of the explicit ones, respectively. The much sharper observability cut on the validated sample than on the corpus is consistent with the per-criterion accuracies reported in Section~\ref{sec:rq1_validation}, where observability is the hardest criterion for the LLM to judge correctly and its errors lean toward over-inclusion.

\begin{table}[t]
\caption{Instruction funnel across the three coding criteria, for the full corpus and for the validated sample. The corpus column counts raw candidates extracted by the LLM. The sample column counts only the 2{,}114 instructions confirmed as genuine during manual validation (2{,}113 true positives plus 1 instruction the LLM missed); this excludes 82 false positives that the LLM extracted but the human annotator rejected.}
\label{tab:rq1_funnel}
\centering
\footnotesize
\rowcolors{1}{gray!15}{white}
\begin{tabular}{lrr}
\rowcolor{black}
{\color{white}Stage} & {\color{white}Sample, confirmed ($N=2{,}114$)} & {\color{white}Corpus, extracted ($N=14{,}902$)} \\
\hline
Total                   & 2{,}114 (100.0\%)      & 14{,}902 (100.0\%) \\
Explicit                & 1{,}403 (66.4\%)       & 10{,}324 (69.3\%) \\
\quad + Actionable      & 1{,}145 (54.2\%)       & 8{,}726 (58.6\%) \\
\quad\quad + Observable & 585 (27.7\%)           & 6{,}552 (44.0\%) \\
\hline
\end{tabular}
\end{table}

Table~\ref{tab:rq1_distribution} reports the corpus-wide category distribution at two funnel stages: explicit-and-actionable ($N=8{,}726$) and the final, fully-qualified dataset ($N=6{,}552$). It also reports a commit-weighted view of the final dataset, which scales each category's count by the number of commits that reuse the snapshot it appears in ($N=106{,}923$ weighted occurrences across the 10{,}763 commits reusing the corpus's snapshots; this weighting uses all commits linked to the snapshots, before the file-level filtering of Section~\ref{sec:data_collection} that yields the 6{,}541 analysis commits). \textit{Implementation Details} is the largest category at every stage and grows further once observability is applied, from 42.8\% of explicit-and-actionable instructions to 50.0\% of the final dataset. Together with \textit{Testing} and \textit{Documentation}, these three categories make up 76.6\% of the final dataset, and this concentration is even sharper (83.2\%) in the commit-weighted view. Other categories shrink more once observability is applied: \textit{Build and Run} drops from 8.8\% of explicit-and-actionable instructions to 2.9\% of the final dataset, and \textit{Build} from 3.8\% to 2.5\%. This shows that process and execution guidance in these categories is often explicit and actionable, but rarely verifiable from the artifacts it leaves behind. For example, ``run the linter'' is a clear, concrete instruction, but nothing in the produced files shows whether the agent actually ran it. The agent distribution stays fairly stable across funnel stages: Copilot accounts for 55.4\% of explicit-and-actionable instructions and 56.9\% of the final dataset, Claude Code for 38.7\% and 37.4\%, and OpenAI Codex for 5.9\% and 5.7\%, roughly matching each agent's share of ACF snapshots in the corpus.

\begin{table}[t]
\caption{Corpus-wide category distribution at the explicit-and-actionable and final (explicit, actionable, and observable) funnel stages, with a commit-weighted view of the final stage. All 19 categories are listed, matching the level of detail used for the category-level statistical tests in \textbf{RQ\textsubscript{3}} (Section~\ref{sec:rq3_compliance}).}
\label{tab:rq1_distribution}
\centering
\footnotesize
\rowcolors{1}{gray!15}{white}
\begin{tabular}{lrrr}
\rowcolor{black}
{\color{white}Category} & {\color{white}Expl.+Act.} & {\color{white}Final} & {\color{white}Weighted} \\
\hline
Implementation Details          & 3{,}738 (42.8\%) & 3{,}279 (50.0\%) & 58{,}944 (55.1\%) \\
Testing                          & 1{,}231 (14.1\%) & 927 (14.1\%)     & 14{,}943 (14.0\%) \\
Documentation                    & 952 (10.9\%)     & 810 (12.4\%)     & 15{,}086 (14.1\%) \\
Build and Run                    & 767 (8.8\%)      & 189 (2.9\%)      & 1{,}772 (1.7\%) \\
Config. and Environment          & 484 (5.5\%)      & 356 (5.4\%)      & 5{,}698 (5.3\%) \\
Development Process              & 373 (4.3\%)      & 221 (3.4\%)      & 2{,}352 (2.2\%) \\
Build                            & 328 (3.8\%)      & 164 (2.5\%)      & 1{,}975 (1.8\%) \\
Quality                          & 222 (2.5\%)      & 133 (2.0\%)      & 1{,}395 (1.3\%) \\
Architecture                     & 152 (1.7\%)      & 131 (2.0\%)      & 1{,}148 (1.1\%) \\
Security                         & 118 (1.4\%)      & 99 (1.5\%)       & 610 (0.6\%) \\
Management                       & 104 (1.2\%)      & 65 (1.0\%)       & 979 (0.9\%) \\
Maintenance                      & 89 (1.0\%)       & 66 (1.0\%)       & 996 (0.9\%) \\
Debugging                        & 46 (0.5\%)       & 36 (0.5\%)       & 333 (0.3\%) \\
Performance                      & 37 (0.4\%)       & 25 (0.4\%)       & 155 (0.1\%) \\
Project Management               & 28 (0.3\%)       & 21 (0.3\%)       & 162 (0.2\%) \\
UI/UX                            & 21 (0.2\%)       & 14 (0.2\%)       & 248 (0.2\%) \\
DevOps                           & 14 (0.2\%)       & 4 (0.1\%)        & 18 (0.0\%) \\
General                          & 13 (0.1\%)       & 11 (0.2\%)       & 88 (0.1\%) \\
AI Integration                   & 9 (0.1\%)        & 1 (0.0\%)        & 21 (0.0\%) \\
\hline
Total                            & 8{,}726          & 6{,}552          & 106{,}923 \\
\hline
\end{tabular}
\end{table}

\subsubsection{Reliability of the Automated Classification}
\label{sec:rq1_validation}

To check how reliable this classification is, we manually validated a stratified sample of 84 context snapshots (13.5\% of the population), covering 2{,}196 individually scored instruction-level decisions. This is the 2{,}202 raw candidates the LLM extracted from these snapshots (Section~\ref{sec:rq1}), plus the 1 instruction it missed, minus 7 candidates for which no decision was recorded. Table~\ref{tab:rq1_extraction} reports extraction precision, recall, and F1-score, overall and by agent. The LLM annotator correctly identified 2{,}113 genuine instructions (TP), incorrectly extracted 82 non-instructions (FP), and missed only 1 instruction that the human annotator later found during full-snapshot review (FN). This gives an overall precision of 96.3\%, recall of 99.9\%\footnote{Recall is computed from a single recorded omission across the entire sample, so while the point estimate is high, it rests on very few observed misses.}, and F1-score of 98.1\%. Performance is consistent across agents: Copilot has the highest F1-score (99.0\%) and OpenAI Codex the lowest (96.0\%), the latter computed over a smaller group of 5 snapshots.

\begin{table}[t]
\caption{Extraction-level performance of the LLM-assisted instruction extraction, computed on the validated sample, overall and by agent.}
\label{tab:rq1_extraction}
\centering
\footnotesize
\rowcolors{1}{gray!15}{white}
\begin{tabular}{lrrrrrrr}
\rowcolor{black}
{\color{white}Group} & {\color{white}Snap.} & {\color{white}TP} & {\color{white}FP} & {\color{white}FN} & {\color{white}Prec.} & {\color{white}Rec.} & {\color{white}F1} \\
\hline
Overall       & 84 & 2{,}113 & 82 & 1 & 0.963 & 0.999 & 0.981 \\
Copilot       & 50 & 1{,}210 & 23 & 1 & 0.981 & 0.999 & 0.990 \\
Claude Code   & 29 & 784     & 49 & 0 & 0.941 & 1.000 & 0.970 \\
OpenAI Codex  & 5  & 119     & 10 & 0 & 0.922 & 1.000 & 0.960 \\
\hline
\end{tabular}
\end{table}

Beyond the extraction itself, Table~\ref{tab:rq1_labels} reports label-level accuracy for each of the six coding decisions, computed over the 2{,}113 confirmed genuine instructions. This covers explicitness, actionability, and observability, alongside inclusion, category, and the candidate observable behavior descriptor. Accuracy is high and consistent for explicitness (95.5\%), actionability (97.3\%), category assignment (96.6\%), and the candidate observable behavior descriptor (97.5\%). The last of these is the field reused as the specification input for measurement script generation in \textbf{RQ\textsubscript{2}}, so its high accuracy supports using it as a reliable input there. Observability (85.0\%) and the resulting inclusion decision (85.2\%) are noticeably less accurate. This confirms that telling artifact-verifiable guidance apart from process-oriented or vague statements is the hardest of the three coding decisions for the LLM annotator, matching the large share lost by process and execution categories at the observability stage of the funnel (Table~\ref{tab:rq1_distribution}). Over 90\% of the disagreements on these two fields corrected an LLM \emph{yes}/\emph{include} decision to \emph{no}/\emph{exclude}, so the annotator's errors lean toward over-inclusion rather than going in a random direction. This means the corpus-wide counts at the observable/final funnel stage (Table~\ref{tab:rq1_funnel}) should be read as upper bounds. OpenAI Codex shows lower category accuracy (87.4\% versus 96.7--98.0\% for the other two agents), again over a small group of snapshots.

\begin{table}[t]
\caption{Label-level accuracy of the LLM's per-criterion coding decisions on confirmed genuine instructions ($N=2{,}113$), overall and by agent.}
\label{tab:rq1_labels}
\centering
\footnotesize
\rowcolors{1}{gray!15}{white}
\begin{tabular}{lrrrr}
\rowcolor{black}
{\color{white}Criterion} & {\color{white}Overall} & {\color{white}Copilot} & {\color{white}Claude Code} & {\color{white}Codex} \\
\hline
Explicitness         & 0.955 & 0.952 & 0.962 & 0.941 \\
Actionability         & 0.973 & 0.967 & 0.983 & 0.975 \\
Observability          & 0.850 & 0.859 & 0.841 & 0.824 \\
Inclusion             & 0.852 & 0.855 & 0.839 & 0.908 \\
Category              & 0.966 & 0.967 & 0.980 & 0.874 \\
Cand. behavior         & 0.975 & 0.988 & 0.954 & 0.983 \\
\hline
\end{tabular}
\end{table}

Overall, the validation shows that the automated classification closely reproduces the explicitness and actionability criteria set during manual calibration (Section~\ref{sec:rq1}), and reproduces observability less accurately and with a systematic lean toward over-inclusion. This gives us enough confidence to treat the LLM-classified corpus as the basis for the \textbf{RQ\textsubscript{1}} dataset, keeping in mind the upper-bound caveat on the observable/final stage noted above.

\bigskip

\smallskip
\steSummaryBox{\faExclamationCircle \hspace{0.05cm} Answer to RQ\textsubscript{1}.}{Applied to the full corpus of 621 context snapshots, the classification pipeline identifies 10{,}324 explicit instructions (69.3\% of all 14{,}902 candidates), of which 8{,}726 (58.6\% of candidates) are additionally actionable, and of which 6{,}552 (44.0\% of candidates, an upper-bound estimate per the validation above) are additionally observable. This fully-qualified dataset spans 95.3\% of all snapshots. Among explicit instructions, observability is the most selective criterion, and it cuts process- and execution-oriented categories hardest (\textit{Build and Run} falls from 8.8\% of explicit-and-actionable instructions to 2.9\% of the final dataset), so that \textit{Implementation Details}, \textit{Testing}, and \textit{Documentation} together make up over three-quarters of the final dataset. Manual validation on a representative sample shows the extraction is reliable (96.3\% precision, 99.9\% recall) and the explicitness, actionability, category, and candidate-behavior labels are accurate (95.5--97.5\%), while observability and the resulting inclusion decision are the least reliable (85.0--85.2\% accuracy, biased toward over-inclusion).}

\subsection{\textbf{RQ\textsubscript{2}} --- To What Extent Can Observable Measurement Proxies Be Operationalized for Explicit and Actionable Instructions Documented in Agent Context Files?} \label{sec:rq2-results}
 
\subsubsection{Operationalization Outcomes} \label{sec:rq2_outcomes}
 
Table~\ref{tab:rq2_operationalization} reports the outcome of the script generation pipeline from Section~\ref{sec:rq2}, applied to the 6{,}552 instructions \textbf{RQ\textsubscript{1}} classified as explicit, actionable, and observable at the corpus level. This is the same final-stage population reported in Table~\ref{tab:rq1_distribution}. Of these, 4{,}123 (62.9\%) reach an accepted measurement script under the sampling budget and corroboration criteria fixed in Section~\ref{sec:rq2}. The remaining 2{,}429 (37.1\%) resist operationalization within that budget and are excluded from \textbf{RQ\textsubscript{3}}.
 
\smallskip \textbf{Category-Level Variation.} Operationalization rate is not the same across instruction categories, and the pattern only partly matches what a simple superficial-versus-substantive split would predict. Among the twelve categories with at least 50 instructions, \textit{Configuration and Environment} operationalizes at the highest rate (269 out of 356, 75.6\%), followed by \textit{Maintenance} (48 out of 66, 72.7\%), \textit{Architecture} (95 out of 131, 72.5\%), \textit{Build and Run} (130 out of 189, 68.8\%), and \textit{Testing} (626 out of 927, 67.5\%), all above the corpus-wide average. This runs counter to a concern raised when this pipeline was designed, that automated operationalization would systematically favor superficial formatting conventions over substantive engineering categories: \textit{Architecture} and \textit{Testing} instructions are, in practice, no harder to operationalize than average, and in this sample somewhat easier. At the opposite end, \textit{Management} (31 out of 65, 47.7\%) and \textit{Security} (48 out of 99, 48.5\%) operationalize at the lowest rates among these categories, more than 14 percentage points below the average. \textit{Implementation Details}, the largest single category by volume (3{,}279 instructions, 50.0\% of the final-stage corpus), operationalizes close to the overall average at 61.2\% (2{,}008 out of 3{,}279). Because it is so large, it dominates the aggregate rate regardless of how other categories behave.
 
\smallskip \textbf{Agent-Level Consistency.} Operationalization rate varies little by coding agent: 2{,}367 out of 3{,}729 (63.5\%) for Copilot, 1{,}520 out of 2{,}449 (62.1\%) for Claude~Code, and 236 out of 374 (63.1\%) for OpenAI~Codex, a spread of only 1.4 percentage points. Which agent's context file an instruction came from has essentially no bearing on whether it can be operationalized. The category-level variation above is the main source of difference in coverage, not the originating agent.
 
\begin{table}[t]
\centering
\caption{Operationalization outcome by instruction category. \textit{Total} reproduces the final-stage, explicit+actionable+observable counts from Table~\ref{tab:rq1_distribution}, listing all 19 categories.}
\label{tab:rq2_operationalization}
\rowcolors{1}{gray!15}{white}
\begin{tabular}{lrrr}
\rowcolor{black}
{\color{white}\textbf{Category}} & {\color{white}\textbf{Total}} & {\color{white}\textbf{Accepted}} & {\color{white}\textbf{Rate}} \\
Implementation Details            & 3{,}279 & 2{,}008 & 61.2\% \\
Testing                           & 927     & 626     & 67.5\% \\
Documentation                     & 810     & 505     & 62.3\% \\
Configuration and Environment     & 356     & 269     & 75.6\% \\
Development Process               & 221     & 130     & 58.8\% \\
Build and Run                     & 189     & 130     & 68.8\% \\
Build                             & 164     & 95      & 57.9\% \\
Quality                           & 133     & 76      & 57.1\% \\
Architecture                      & 131     & 95      & 72.5\% \\
Security                          & 99      & 48      & 48.5\% \\
Maintenance                       & 66      & 48      & 72.7\% \\
Management                        & 65      & 31      & 47.7\% \\
Debugging                         & 36      & 19      & 52.8\% \\
Performance                       & 25      & 15      & 60.0\% \\
Project Management                & 21      & 8       & 38.1\% \\
UI/UX                             & 14      & 7       & 50.0\% \\
General                           & 11      & 10      & 90.9\% \\
DevOps                            & 4       & 3       & 75.0\% \\
AI Integration                    & 1       & 0       & 0.0\% \\\hline
\textbf{Total}                    & \textbf{6{,}552} & \textbf{4{,}123} & \textbf{62.9\%} \\\hline
\end{tabular}
\end{table}
 
\subsubsection{Exclusion Patterns and Generation Reliability} \label{sec:rq2_diagnostics}
 
\revised{Table~\ref{tab:rq2_exclusions} breaks down the 2{,}429 excluded instructions by the bar they failed. Most of them (1{,}392, 57.3\%) fail the minimum-correctness bar: for 1{,}006 (41.4\%) no generated script passed even one fixture, and for 386 (15.9\%) the best script passed only one. A further 997 (41.1\%) clear the correctness bar, but no independently generated script agrees with the best one. Only 40 (1.6\%) fail at candidate generation, with fewer than two syntactically valid scripts.}
 
This shows that generation-stage reliability is high, and exclusion is mostly a correctness or agreement problem, not a formatting or parsing one. Across all 6{,}552 instructions, the pipeline obtained the full requested budget of ten candidate scripts for 6{,}223 (95.0\%) and the full ten candidate fixtures for 5{,}786 (88.3\%). Truncated or malformed candidates are the exception, not the rule. Among the 4{,}123 accepted instructions, the winning consensus group has a mean size of 6.3 out of 10, and this splits into two clusters: 825 instructions (20.0\% of accepted) reach full unanimous agreement across all ten sampled scripts, while 589 (14.3\%) are accepted on the minimum admissible corroboration of exactly two agreeing scripts. The remaining 65.7\% are spread fairly evenly in between. Operationalizability is therefore not a clean two-way split: many instructions have a candidate observable behavior specific and mechanical enough that independently sampled scripts converge on it readily, but the corroboration threshold is decisive for a substantial share of cases, both among accepted instructions (the 14.3\% accepted on exactly two agreeing scripts) and among excluded ones (the 997 instructions whose correct-looking script found no corroborating peer).

Consistent with the permissive-default tendency noted in Section~\ref{sec:rq3} as the reason for the applicability filter, 2{,}569 of the 4{,}123 accepted scripts (62.3\%) return \texttt{True} with no supporting evidence when checked against a commit that touches no relevant files. The applicability filter is built to exclude exactly these uninformative pairings from the \textbf{RQ\textsubscript{3}} analysis, so this default does not by itself bias the compliance rates reported there. It does mean, though, that any instruction-commit pair that slips past the filter despite being irrelevant will count as compliant rather than as a neutral non-observation. The \textbf{RQ\textsubscript{3}} compliance figures should therefore be read as an optimistic estimate: the reported non-compliance rate is a floor on true non-adherence, not a precise ceiling, since any leakage through the applicability filter can only push apparent compliance up, never down.

The manual spot-check described in Section~\ref{sec:rq2} finds that 67 of the 94 sampled accepted scripts (71.3\%) correctly implement their instruction's candidate observable behavior. Conversely, roughly three in ten accepted scripts do not faithfully implement their specification, so passing both acceptance bars is evidence of, not a guarantee of, a correct operationalization. The disagreements cluster around a small number of recurring patterns rather than spreading evenly. The most common is a script that hardcodes an example from the ACF itself, a specific filename, type name, or argument value used to illustrate a convention, as a literal requirement instead of generalizing the underlying rule. Genuine implementation bugs, such as a condition that can never be true, a missing import, or a regex typo that silently disables half a check, are comparatively rare in the sample. Because these errors are systematic rather than random, they cannot be assumed to cancel out: a hardcoded example typically makes a script stricter than the instruction it implements, which would bias the affected instructions' \textbf{RQ\textsubscript{3}} compliance downward. The \textbf{RQ\textsubscript{3}} rates should therefore be read with this measurement error in mind, alongside the optimistic bias from permissive defaults discussed above.

\begin{table}[t]
\centering
\caption{Exclusion reason for the 2{,}429 instructions that did not reach an accepted script. Percentages may not sum to 100\% due to rounding.}
\label{tab:rq2_exclusions}
\rowcolors{1}{gray!15}{white}
\begin{tabular}{lrr}
\rowcolor{black}
{\color{white}\textbf{Exclusion Reason}} & {\color{white}\textbf{Count}} & {\color{white}\textbf{\% of Exclusions}} \\
No script passed any fixture      & 1{,}006 & 41.4\% \\
No corroborating script           & 997     & 41.0\% \\
Top script below min.\ passed     & 386     & 15.9\% \\
Insufficient valid candidates     & 40      & 1.6\% \\\hline
\textbf{Total excluded}           & \textbf{2{,}429} & \textbf{100.0\%} \\\hline
\end{tabular}
\end{table}

\smallskip
\steSummaryBox{\faExclamationCircle \hspace{0.05cm} Answer to RQ\textsubscript{2}.}{A majority, 4{,}123 out of 6{,}552 (62.9\%), of the explicit, actionable, and observable instructions identified in \textbf{RQ\textsubscript{1}} can be operationalized as executable measurement scripts through test-driven, consensus-based generation, although a manual spot-check confirms that only 71.3\% of accepted scripts faithfully implement their instruction's specification. Operationalization rate is essentially unaffected by which coding agent's context file an instruction came from, but varies meaningfully by category: \textit{Configuration and Environment}, \textit{Architecture}, and \textit{Testing} instructions operationalize above the corpus average, while \textit{Management} (47.7\%) and \textit{Security} (48.5\%) instructions operationalize at markedly lower rates. \revised{Most exclusions (57.3\%) happen because no generated script is correct enough, and 41.1\% because no second script agrees with the best one. Almost none (1.6\%) come from missing or malformed candidates.} Because the operationalized subset is not a uniform sample of \textbf{RQ\textsubscript{1}}'s instruction population, particularly for \textit{Security} and \textit{Management}, the adherence results reported for \textbf{RQ\textsubscript{3}} describe compliance within this operationalizable subset, not the full space of explicit and actionable directives documented in ACFs.}

\subsection{\textbf{RQ\textsubscript{3}} --- To What Extent Are Observable Instructions Documented in Agent Context Files Reflected in Downstream Agent-Generated Commits?} \label{sec:rq3-results}

\subsubsection{Evaluation Coverage} \label{sec:rq3_coverage}

Table~\ref{tab:rq3_coverage} reports how many of the 4{,}123 instructions operationalized in \textbf{RQ\textsubscript{2}} could actually be checked against a real commit. An instruction contributes an adherence observation only if at least one commit generated under its originating context snapshot touches a file matching that instruction's own applicability pattern, derived in Section~\ref{sec:rq3} from the instruction's accepted fixtures. Of the 4{,}123 operationalized instructions, 2{,}053 (49.8\%) have at least one applicable commit. The remaining 2{,}070 (50.2\%) are operationalizable in principle but never meet a commit that touches the relevant file type in the mined dataset, and are excluded from the compliance analysis on that basis. This is a further loss beyond \textbf{RQ\textsubscript{2}}'s own operationalization funnel: across the full pipeline, only 2{,}053 of the original 6{,}552 explicit, actionable, and observable instructions from \textbf{RQ\textsubscript{1}} (31.3\%) end up producing an adherence observation. The 2{,}053 evaluated instructions are checked against 18{,}594 instruction-commit pairs in total.

\smallskip \textbf{Category-Level Coverage.} Coverage is uneven across categories and does not track the operationalization rate from \textbf{RQ\textsubscript{2}}. Among the nine categories with at least 50 operationalized instructions, \textit{Documentation} reaches the highest coverage (340 out of 505, 67.3\%), while \textit{Configuration and Environment} (99 out of 269, 36.8\%) and \textit{Development Process} (50 out of 130, 38.5\%) fall well below the corpus-wide average of 49.8\%; among smaller categories, \textit{Maintenance} (15 out of 48, 31.2\%) and \textit{Security} (18 out of 48, 37.5\%) are similarly low. This happens even though \textit{Configuration and Environment} had the highest operationalization rate of any category in \textbf{RQ\textsubscript{2}} (Table~\ref{tab:rq2_operationalization}). This gap shows that being operationalizable and being evaluable are governed by different things: an instruction can be easy to write a correct measurement script for, and still rarely meet a matching commit in the mined dataset, because the file types it concerns are touched infrequently relative to how often the convention could in principle be checked.

\smallskip \textbf{Agent-Level Coverage.} Coverage is close to uniform by originating agent: 799 out of 1{,}520 (52.6\%) for Claude~Code, 1{,}137 out of 2{,}367 (48.0\%) for Copilot, and 117 out of 236 (49.6\%) for OpenAI~Codex, a spread of 4.6 percentage points. As with \textbf{RQ\textsubscript{2}}'s operationalization rate, which agent's context file an instruction came from has little bearing on whether it is ever evaluable. Category composition, not agent, is again the main source of variation.

\begin{table}[t]
\centering
\caption{Evaluation coverage and compliance outcome by instruction category. \textit{Operationalized} reproduces the \textbf{RQ\textsubscript{2}} accepted counts from Table~\ref{tab:rq2_operationalization}. \textit{Evaluated} is the number of operationalized instructions with at least one applicable commit; \textit{Mean}/\textit{Median} are computed over the per-instruction compliance rate among evaluated instructions only. \textit{AI Integration} is omitted from the row listing (0 operationalized instructions in Table~\ref{tab:rq2_operationalization}, hence 0 evaluated) but is included in the Total row for completeness; it also does not contribute a group to the category-level significance test in Section~\ref{sec:rq3_compliance}, which is accordingly computed over 18 categories.}
\label{tab:rq3_coverage}
\rowcolors{1}{gray!15}{white}
\begin{tabular}{lrrrrr}
\rowcolor{black}
{\color{white}\textbf{Category}} & {\color{white}\textbf{Operat.}} & {\color{white}\textbf{Eval.}} & {\color{white}\textbf{Cov.}} & {\color{white}\textbf{Comp. Mean}} & {\color{white}\textbf{Comp. Median}} \\
Implementation Details            & 2{,}008 & 1{,}022 & 50.9\% & 63.6\% & 100.0\% \\
Testing                           & 626     & 270     & 43.1\% & 49.2\% & 37.5\%  \\
Documentation                     & 505     & 340     & 67.3\% & 60.2\% & 100.0\% \\
Configuration and Environment     & 269     & 99      & 36.8\% & 56.3\% & 100.0\% \\
Build and Run                     & 130     & 62      & 47.7\% & 69.9\% & 100.0\% \\
Development Process               & 130     & 50      & 38.5\% & 34.0\% & 0.0\%   \\
Architecture                      & 95      & 40      & 42.1\% & 38.2\% & 0.0\%   \\
Build                             & 95      & 48      & 50.5\% & 67.5\% & 100.0\% \\
Quality                           & 76      & 37      & 48.7\% & 63.5\% & 100.0\% \\
Maintenance                       & 48      & 15      & 31.2\% & 85.3\% & 100.0\% \\
Security                          & 48      & 18      & 37.5\% & 58.6\% & 98.7\%  \\
Management                        & 31      & 24      & 77.4\% & 44.6\% & 0.0\%   \\
Debugging                         & 19      & 4       & 21.1\% & 21.9\% & 0.0\%   \\
Performance                       & 15      & 9       & 60.0\% & 65.1\% & 100.0\% \\
General                           & 10      & 7       & 70.0\% & 71.4\% & 100.0\% \\
Project Management                & 8       & 3       & 37.5\% & 93.3\% & 100.0\% \\
UI/UX                             & 7       & 3       & 42.9\% & 91.7\% & 100.0\% \\
DevOps                            & 3       & 2       & 66.7\% & 50.0\% & 50.0\%  \\\hline
\textbf{Total}                    & \textbf{4{,}123} & \textbf{2{,}053} & \textbf{49.8\%} & \textbf{59.8\%} & \textbf{100.0\%} \\\hline
\end{tabular}
\end{table}

\subsubsection{Compliance Outcomes} \label{sec:rq3_compliance}

Across the 2{,}053 evaluated instructions, the mean per-instruction compliance rate is 59.8\%, but the median is 100.0\%. This gap between mean and median points to a strongly bimodal distribution rather than a spread of partial compliance: 1{,}116 instructions (54.4\%) are followed in every applicable commit checked, 766 (37.3\%) are followed in none, and only 171 (8.3\%) fall strictly in between. Part of this pattern comes from how thin the evidence often is: the median instruction has only 3 applicable commits, and 540 instructions (26.3\%) have exactly one, so many of the 0\% or 100\% rates rest on a single observation. This does not bias the group comparisons that follow, since those operate on per-instruction rates precisely to avoid counting the same evidence twice (Section~\ref{sec:rq3-stats}). But it does mean individual instruction-level rates should be read as rough signals, not precise estimates.

\smallskip \textbf{Category-Level Variation.} A Kruskal-Wallis test across the 18 instruction categories with at least one evaluated instruction finds a significant difference between them ($H = 59.13$, $p < .001$); \textit{AI Integration} contributes no group to this test, since it has zero operationalized instructions in \textbf{RQ\textsubscript{2}} (Table~\ref{tab:rq2_operationalization}) and zero evaluated instructions here (Table~\ref{tab:rq3_coverage}). Post-hoc pairwise Mann-Whitney U tests with Bonferroni correction across the 153 category pairs identify four pairs that remain significant. Three involve \textit{Development Process} as the lower-compliance side; the fourth compares the two largest categories of the final dataset, \textit{Implementation Details} and \textit{Testing}. Effect sizes are reported as Cliff's $\delta$ of the first-named group against the second. \textit{Development Process} instructions (mean 34.0\%, median 0.0\%) comply significantly less often than \textit{Build and Run} ($p_{\text{Bonf}} = .013$, $\delta = -0.39$), \textit{Documentation} ($p_{\text{Bonf}} = .037$, $\delta = -0.29$), and \textit{Implementation Details} ($p_{\text{Bonf}} = .002$, $\delta = -0.32$), each a small-to-medium effect. \textit{Implementation Details} also complies significantly more often than \textit{Testing} ($p_{\text{Bonf}} = .002$, $\delta = 0.15$), a small effect. \textit{Architecture} (mean 38.2\%, median 0.0\%) and \textit{Debugging} (mean 21.9\%, median 0.0\%) show similarly low compliance to \textit{Development Process} on paper, but neither survives Bonferroni correction against any other category: \textit{Architecture} because so many comparisons were tested, \textit{Debugging} because its support (4 evaluated instructions) is too small to carry statistical power. We report both as descriptive only. Taken together, the surviving comparisons single out \textit{Development Process} as the one category with a significant compliance gap relative to several others in the full sample, rather than pointing to broad, category-driven variation across the whole corpus; Section~\ref{sec:rq3_sensitivity} shows that this gap, unlike the \textit{Implementation Details}-versus-\textit{Testing} difference, does not survive once thin-evidence instructions are excluded.

\smallskip \textbf{Agent-Level Variation.} Table~\ref{tab:rq3_agent} reports compliance by originating agent. A Kruskal-Wallis test across the three agents finds a significant difference between them ($H = 41.58$, $p < .001$). Copilot instructions comply at a mean rate of 66.0\% (median 100.0\%), significantly higher than both Claude~Code (52.2\% mean, $p_{\text{Bonf}} < .001$, $\delta = 0.15$, a small effect) and OpenAI~Codex (51.2\% mean, $p_{\text{Bonf}} = .018$, $\delta = 0.13$, a negligible-to-small effect). Claude~Code and OpenAI~Codex are statistically indistinguishable from each other ($p_{\text{Bonf}} = 1.0$, $\delta = -0.0005$). So the finding is not that three agents diverge from one another in three directions, but that Copilot-authored commits follow their own context file's instructions somewhat more often than Claude~Code's or Codex's do, while the latter two are, on this measure, identical. OpenAI~Codex's estimate rests on the smallest sample of the three (117 evaluated instructions against 799 and 1{,}137 for the other two agents), so while the Copilot-versus-Codex difference reaches significance, it should be read with that imbalance in mind.

\subsubsection{Sensitivity to Applicable-Commit Thresholds} \label{sec:rq3_sensitivity}

\revised{Because 540 of the 2{,}053 evaluated instructions (26.3\%) have exactly one applicable commit, which forces their individual compliance rate to 0\% or 100\% by construction, we check whether the results above hold up once these thin-evidence instructions are excluded. We repeat the bimodal-split computation and both Kruskal-Wallis analyses restricted to instructions with at least 3 (n = 1{,}144) and at least 5 (n = 819) applicable commits.}

\revised{The bimodal shape itself holds up well. The share of instructions at a clean extreme, 0\% or 100\%, drops only a little as the thin-evidence tail is excluded: from 91.7\% in the full sample to 87.3\% at the $\geq 3$ threshold and 86.2\% at $\geq 5$. The share falling strictly in between rises to match, from 8.3\% to 12.7\% to 13.8\%. Even at the strictest threshold, most instructions are still either always or never followed, not spread across a range.}

\revised{The category-level difference across groups also stays significant at both thresholds ($H = 40.79$, $p < .001$ at $\geq 3$; $H = 43.97$, $p < .001$ at $\geq 5$), but the specific pairs behind it do not all hold up. None of \textit{Development Process}'s three significant comparisons in the full sample (versus \textit{Build and Run}, \textit{Documentation}, \textit{Implementation Details}) survive Bonferroni correction at either threshold, even though its mean compliance stays the lowest among categories with enough support at each threshold (34.0\% at $n=50$, 25.0\% at $n=14$, 21.4\% at $n=7$). Only the \textit{Implementation Details}-versus-\textit{Testing} comparison stays significant at every threshold; this is the one category-level finding that fully holds up under this check. Since \textit{Development Process}'s own evaluated sample shrinks from 50 to 14 to 7 instructions as we filter, this looks like a loss of statistical power on an already small group, not a reversal of the underlying gap.}

\revised{The agent-level result splits the same way. The Copilot-versus-Claude~Code gap stays significant at every threshold ($p_{\text{Bonf}} < .001$, $.002$, $.004$ at the full, $\geq 3$, and $\geq 5$ samples, $\delta$ consistently between $0.12$ and $0.15$), so this is the one agent-level comparison that fully holds up. The Copilot-versus-OpenAI~Codex gap, by contrast, is significant only in the full sample ($p_{\text{Bonf}} = .018$) and not at $\geq 3$ ($p_{\text{Bonf}} = .165$) or $\geq 5$ ($p_{\text{Bonf}} = .102$), even though the direction of the effect is unchanged and its magnitude grows ($\delta = 0.13$, $0.14$, $0.23$) as Codex's already small evaluated sample shrinks further, from 117 to 56 to 22 instructions, under filtering. Claude~Code and OpenAI~Codex remain statistically indistinguishable at every threshold, matching the full-sample result.}

\revised{Taken together, this check supports the paper's core compliance finding, the bimodal split and the Copilot-versus-Claude~Code gap, as holding up once thin-evidence instructions are excluded. It also shows that \textit{Development Process}'s specific compliance gap and Copilot's specific edge over OpenAI~Codex point in the same direction but are not statistically solid at stricter evidence thresholds, and we read them accordingly in what follows.}

\begin{table}[t]
\centering
\caption{Compliance outcome by coding agent. \textit{Operationalized} and \textit{Evaluated} counts as in Table~\ref{tab:rq3_coverage}.}
\label{tab:rq3_agent}
\rowcolors{1}{gray!15}{white}
\begin{tabular}{lrrrrr}
\rowcolor{black}
{\color{white}\textbf{Agent}} & {\color{white}\textbf{Operat.}} & {\color{white}\textbf{Eval.}} & {\color{white}\textbf{Cov.}} & {\color{white}\textbf{Mean}} & {\color{white}\textbf{Median}} \\
Copilot                            & 2{,}367 & 1{,}137 & 48.0\% & 66.0\% & 100.0\% \\
Claude Code                        & 1{,}520 & 799     & 52.6\% & 52.2\% & 66.7\%  \\
OpenAI Codex                       & 236     & 117     & 49.6\% & 51.2\% & 100.0\% \\\hline
\textbf{Total}                     & \textbf{4{,}123} & \textbf{2{,}053} & \textbf{49.8\%} & \textbf{59.8\%} & \textbf{100.0\%} \\\hline
\end{tabular}
\end{table}

\smallskip
\steSummaryBox{\faExclamationCircle \hspace{0.05cm} Answer to RQ\textsubscript{3}.}{Of the 4{,}123 instructions operationalized in \textbf{RQ\textsubscript{2}}, 2{,}053 (49.8\%) are ever checked against a real commit, yielding 18{,}594 instruction--commit observations; the remainder are operationalizable but never meet an applicable commit in the mined dataset. Among evaluated instructions, mean compliance is 59.8\%, but the distribution is strongly bimodal: 54.4\% of instructions are followed in every applicable commit and 37.3\% in none, rather than spread across a range of partial adherence. Compliance varies significantly by category ($p<.001$), but few pairwise differences survive correction for multiple comparisons. \textit{Implementation Details} instructions are followed more often than \textit{Testing} ones, the only category-level difference that holds at every evidence threshold. \textit{Development Process} instructions \revised{are the least compliant among categories with enough support and differ significantly from three others in the full sample, though this gap is not solid under a sensitivity check that restricts the comparison to instructions with more applicable-commit evidence (Section~\ref{sec:rq3_sensitivity})}. Compliance also varies significantly by agent ($p<.001$): Copilot-authored commits comply significantly more often than Claude~Code's\revised{, a gap that holds up under the same sensitivity check, while Copilot's edge over OpenAI~Codex specifically does not}; Claude~Code and OpenAI~Codex are statistically indistinguishable from one another\revised{ at every threshold tested}. Overall, explicit and actionable ACF instructions are reflected in agent behavior more often than not, but adherence is neither uniform across instruction categories nor across coding agents.}
\section{Discussion and Implications}
In this section, we discuss our findings further and draw practical implications for practitioners using autonomous coding agents in their workflow.

To the best of our knowledge, this is among the first articles to evaluate ACF governance at the level of the single instruction, rather than the whole file. Prior studies describe ACF content through high-level categories~\cite{chatlatanagulchai2025agent, chatlatanagulchai2025use}, describe large differences in structure across projects~\cite{mohsenimofidi2025context}, or treat the ACF as an all-or-nothing treatment, measuring the effect of its presence rather than of what it says~\cite{lulla2026impactagentsmdfilesefficiency}. Our breakdown into explicit, actionable, and observable instructions shows that this broader view hides substantial differences: indeed, within the same category, individual instructions differ sharply in whether they can be verified at all, let alone whether they are followed.
\smallskip

\steDiscussionBox{\faSuitcase \ Researchers studying agent governance should treat the single instruction, not the file or the category, as the unit of analysis. Aggregate compliance claims summarize a strongly bimodal population (Section~\ref{sec:rq3_compliance}), not a stable property of the artifact itself.}

\subsection{On Agent Instructions in the Wild}

Of the 14{,}902 candidate instructions extracted from the corpus, 69.3\% are explicit and only 58.6\% are also actionable (Table~\ref{tab:rq1_funnel}). These figures describe how ACF authors write guidance in practice, independent of whether an agent later complies with it. Prompt engineering work argues that instruction clarity and specificity are key to reliable LLM behavior \cite{sasaki2024systematic, della2026toward}: vague or non-actionable guidance puts the burden of interpretation on the model. Read against this, our extraction funnel suggests that \textbf{a large share of ACF content in the wild is written more like documentation for human readers than instruction for an agent}: nearly a third of statements are not explicit, and one in six of the explicit ones does not prescribe a concrete, performable action. This holds regardless of the compliance question we answer in RQ\textsubscript{3}. An instruction can still be followed well by an agent that is capable of resolving its ambiguity, but low explicitness or actionability is still an authoring weakness. It limits how reliably an instruction steers behavior across different agents and models.

Observability is a separate concern from explicitness and actionability. An instruction can be perfectly explicit and actionable to carry out, like ``run the linter before committing,'' while leaving no trace in the artifacts an outside observer can check. This matters for building automated compliance mechanisms, since only instructions that leave such a trace can ever be audited without looking at an agent's internal reasoning. Among explicit instructions, observability is the most selective of our criteria, and it hits process-oriented guidance hardest: \textit{Build and Run} instructions drop from 8.8\% of explicit-and-actionable instructions to 2.9\% of the final dataset. The losses continue further down the pipeline: only 62.9\% of observable instructions turn into a working, corroborated script (Section~\ref{sec:rq2_outcomes}), and only 49.8\% of those ever meet an applicable commit (Section~\ref{sec:rq3_coverage}). In the end, just 2{,}053 of the original 14{,}902 candidate instructions (13.8\%) become the basis for an actual, artifact-grounded compliance observation.

\subsection{On Instruction Compliance}

Compliance is strongly bimodal: 54.4\% of the 2{,}053 evaluated instructions are followed in every applicable commit checked, 37.3\% are followed in none, and only 8.3\% fall strictly in between (Section~\ref{sec:rq3_compliance}). Based on our artifact-level proxies, ACF-based governance behaves less like a dial agents turn to a partial setting, and more like a switch. Part of this shape is mechanical: each commit yields a binary verdict and the median instruction is checked against only 3 commits, so intermediate rates are hard to obtain. But the split survives when instructions with fewer than 3 or 5 applicable commits are excluded (Section~\ref{sec:rq3_sensitivity}), so it is not only an artifact of thin evidence. This has a practical meaning: non-compliance is concentrated in instructions that are essentially never followed, rather than spread thinly across all of them. Checking a handful of rules therefore gives little assurance about the rest: a review that happens to sample consistently followed instructions will miss the 37.3\% that are never respected. The picture is also likely optimistic rather than pessimistic: most compliance-checking scripts are permissive, so any leftover leakage through the applicability filter can only push compliance figures up, never down. Reported non-compliance is a floor, not a ceiling.

Non-compliance is also not spread evenly across the corpus. \textit{Development Process} instructions comply significantly less often than \textit{Build and Run}, \textit{Documentation}, and \textit{Implementation Details} (mean 34.0\%, median 0.0\%)\revised{, though this specific gap does not hold up under a sensitivity check that restricts the comparison to instructions with more applicable-commit evidence (Section~\ref{sec:rq3_sensitivity}), so we treat the direction as suggestive rather than firmly established}. This is plausibly because workflow and collaboration conventions, such as commit formatting or branching practices, sit at the edge of what any artifact-based proxy, or the agent itself, reliably treats as a binding rule rather than as background context. At the other end, \textit{Implementation Details} and \textit{Documentation}, two of the three categories that together make up 76.6\% of the final RQ\textsubscript{1} dataset (Table~\ref{tab:rq1_distribution}), are followed in a majority of cases (means of 63.6\% and 60.2\%, medians of 100\%). The third, \textit{Testing}, complies at a noticeably lower rate (mean 49.2\%, median 37.5\%) and significantly less often than \textit{Implementation Details}, the one category-level difference that holds at every evidence threshold (Section~\ref{sec:rq3_sensitivity}). \textbf{ACF-based governance works best for code-level conventions, and less well for testing and process-oriented guidance}.

Agent identity matters too, as a separate axis from category. Copilot-authored commits comply with their own context file at a significantly higher mean rate (66.0\%) than either Claude Code (52.2\%) or OpenAI Codex (51.2\%), which are statistically indistinguishable from each other (Section~\ref{sec:rq3_compliance}). Operationalization and coverage rates are nearly uniform across agents (Sections~\ref{sec:rq2_outcomes}, \ref{sec:rq3_coverage}), so the gap does not stem from which instructions could be measured for each agent. It is \revised{consistent with} real differences in how each agent's tooling loads, prioritizes, or re-asserts ACF content during a session, but our agent-level comparison does not control for the different category mix of each agent's ACFs, so part of the gap may reflect what those ACFs ask for rather than how faithfully each agent follows them. This is consistent with Huang et al.~\cite{huang2025professional}, who find that developers already compensate for this uncertainty through active, ongoing supervision rather than trusting agents by default. \revised{We treat this as an association, not an established cause: category mix, differences in repositories and languages across agents, and the uneven attribution precision across agents noted under external validity, remain plausible contributing factors we cannot fully rule out. Of the two pairwise gaps behind this result, Copilot's edge over Claude~Code holds up under a sensitivity check that restricts the comparison to instructions with more applicable-commit evidence; its edge over OpenAI~Codex specifically does not (Section~\ref{sec:rq3_sensitivity}), and should be weighted accordingly.}

\subsection{\revised{Implications for ACF Design}}

\revised{Together, the extraction-funnel and compliance results speak directly to how ACFs should be written. The gap between explicit and actionable instructions (Section~\ref{sec:rq1-results}) shows that a meaningful share of ACF content is written closer to documentation than to agent-directed instruction. This is an authoring problem with a concrete fix: ACF authors should write for the agent, not the reader. State one concrete behavior per rule, use imperative phrasing, and avoid vague qualifiers such as ``when appropriate'' or ``if needed,'' which put the burden of interpretation on the model instead of resolving it in the file itself.}

\revised{Observability adds a second, mostly separate authoring concern. An instruction can be explicit and actionable and still leave no trace an outside observer, human or automated, can check. Since only observable instructions can be audited without looking at an agent's internal reasoning, developers should prefer instructions that are checkable from the artifacts an agent modifies, such as file properties, naming conventions, or structural rules, over process-only directives whose execution leaves nothing behind. Tool vendors could support this directly at authoring time, flagging non-observable instructions before the gap is discovered later, as our own extraction pipeline does after the fact.}

\revised{The category-level compliance results (Section~\ref{sec:rq3_compliance}) point to a further, more \textbf{tentative} design implication, offered with the caveat that the pairwise comparison behind it is not solid under a sensitivity check (Section~\ref{sec:rq3_sensitivity}). \textit{Development Process} instructions are descriptively both harder to observe and less complied with than \textit{Implementation Details}, \textit{Testing}, or \textit{Documentation}. Workflow-level conventions, such as commit message formatting or branching discipline, are plausibly the kind of guidance that natural-language phrasing in an ACF is least suited to carry reliably. If this pattern holds up under further evidence, authors may be better served by encoding such conventions directly into project tooling, such as commit hooks, branch-protection rules, or CI configuration, than by relying on prose an agent may or may not follow consistently. We flag this as a direction worth checking further, not a settled recommendation.}

\smallskip
\steDiscussionBox{\faSuitcase \ \revised{ACF authors should write one concrete, imperative, artifact-checkable behavior per rule, avoiding vague qualifiers and process-only directives whose execution leaves no trace. Tool vendors should flag non-observable instructions at authoring time. And workflow-level conventions, \textit{Development Process} guidance in particular, though this specific category-level result is descriptive rather than statistically solid (Section~\ref{sec:rq3_sensitivity}), may be better encoded in project tooling than left as ACF prose.}}

\subsection{\revised{Implications for Engineering Agentic Systems}}

\revised{Beyond how ACFs are written, our results speak to how agentic systems should be built to govern the instructions ACFs contain. The bimodal shape of compliance (Section~\ref{sec:rq3_compliance}) means spot-checking is a poor substitute for systematic verification. Because non-compliance concentrates in instructions that are essentially never followed, checking a few rules says little about the others: sampling rules that happen to be respected will miss those that are consistently ignored. This points toward automated, continuously-run compliance checks over occasional review, aimed specifically at the instruction categories and agents our results flag as least reliable, rather than applied evenly across an ACF whose instructions are, as shown above, far from equally reliable.}

\revised{The agent-identity effect strengthens this point beyond day-to-day monitoring: it bears on how verification mechanisms should be designed in the first place. Compliance differences across Copilot, Claude Code, and OpenAI Codex do not stem from differences in what could be operationalized or evaluated for each agent, so reliability of ACF adherence may be better understood as a property of the \textit{ACF and agent together} than of the ACF alone, though, as noted above, we cannot fully rule out category-mix, repository-, or language-level confounds. This reliability should not be assumed to carry over if a team switches to a different agent or adopts a new one. Verification mechanisms built on top of an ACF therefore cannot be agent-agnostic by default. They need to be calibrated, or re-checked, per agent.}

\revised{Our own measurement pipeline (Section~\ref{sec:rq2}) is a concrete example of the kind of mechanism this argues for. Sampling scripts and fixtures independently from each instruction’s behavioral descriptor, and accepting only scripts that are both minimally correct and independently corroborated, is itself a verification layer grounded in artifacts, sitting alongside the ACF's natural-language content rather than inside it. That 62.9\% of observable instructions could be operationalized this way suggests that automated, per-instruction compliance checking is practical at scale for a large share of ACF content today, not only a theoretical possibility. It is not yet hands-off: our spot-check found roughly three in ten accepted scripts implementing their instruction incorrectly (Section~\ref{sec:rq2_diagnostics}), so generated checks still need human review before they are trusted as enforcement.}

\revised{Taken together, these results support the broader point that ACFs are one layer of governance, not sufficient on their own. Natural-language instructions are still needed to communicate intent, but for the categories and agents our findings show to be unreliable, they need to be paired with independent, artifact-grounded enforcement, such as automated compliance checks, CI-integrated verification, or agent-specific safeguards, rather than trusted on their own.}

\smallskip
\steDiscussionBox{\faSuitcase \ \revised{Practitioners should pair ACFs with automated, continuously-run compliance checks aimed at the least reliable categories and agents rather than spot-checking. Verification mechanisms should be calibrated per ACF-and-agent pair rather than assumed to carry over across agents. And our own test-driven script-generation pipeline shows that artifact-grounded, per-instruction verification is practical today for a majority of observable ACF content, provided generated checks are reviewed before being used as enforcement.}}
\section{Threats to Validity} \label{sec:threats}
This section outlines potential threats to the validity of our study and the mitigation strategies implemented.

\textbf{Internal validity} concerns whether observed relationships can be attributed to the studied factors.
A first threat is the absence of a counterfactual: we never observe what an agent would produce in the absence of a given instruction, so compliance may partly reflect ecosystem or model defaults rather than the ACF itself. This is tempered by our explicitness/actionability criteria, which favor project-specific directives over generic best practices, for which alternative explanations are weaker; a full ablation, however, is out of scope. 

A second threat concerns snapshot--commit linking: if a reconstructed snapshot does not match what the agent actually used at generation time, compliance may be misattributed. We mitigate this through content-based deduplication of snapshots and by linking each commit to the snapshot version active when it was generated (Section~\ref{sec:data_collection}), backed by automated completeness checks.

A third threat concerns the agent-level comparison in RQ\textsubscript{3}: per-instruction compliance rates are compared across agents without controlling for instruction category, and the agents' ACFs differ in category mix. Because compliance also varies by category, part of the observed agent gap may reflect what each agent's ACFs ask for rather than how faithfully each agent follows them; together with repository- and language-level differences across agents, this is why we interpret the agent-level result as an association rather than a causal effect of agent identity.

Another threat is residual noise in agent-authorship attribution, which relies on AIDev's heuristic signals~\cite{li2025aiteammates}; we mitigate this by restricting analysis to the \textit{strict agent subset}, AIDev's higher-confidence tier, without claiming a validated precision figure. 

Finally, the LLM components of our pipeline are inherently stochastic, and prior work shows LLMs can produce different code even under low-temperature or greedy decoding~\cite{ouyang2025empirical}; this is especially relevant for RQ\textsubscript{2}, where script and fixture generation deliberately uses a higher temperature (0.8) for candidate diversity. We bound this threat with fixed inference parameters throughout, and, for RQ\textsubscript{2} specifically, by construction: accepted scripts must clear the dual-bar criterion described below, so validity rests on empirical pass/fail evidence rather than on generation determinism.

\textbf{Construct validity} concerns how well measures capture the intended concepts.
A primary threat lies in the acceptance criterion used to select measurement scripts in RQ\textsubscript{2}. We adapt \textsc{CodeT}'s dual execution agreement from exact-set consensus to a dual-bar criterion, consisting of a minimum-correctness bar and a 60\% overlap-based corroboration bar, because a pilot showed that exact agreement essentially never occurs at our sampling scale. Both thresholds were fixed from that same pilot rather than calibrated independently\revised{. In addition, an accepted script reflects only one valid operationalization of an instruction's behavioral descriptor, so instructions that can be satisfied through several legitimate realizations may be undercounted, whereas the vacuous-compliance risk discussed below overcounts them}. We partially mitigate this by manually spot-checking a statistically sized, category-stratified sample of accepted scripts (Section~\ref{sec:rq2}) and reporting the agreement rate between this check and the automated decision as a direct reliability measure. That check, however, also quantifies the residual threat: 67 of 94 sampled scripts (71.3\%) correctly implement their specification, so roughly three in ten accepted scripts measure something other than what the instruction prescribes. The most frequent error, hardcoding an example from the ACF as a literal requirement, makes a script stricter than its instruction and would bias the affected compliance rates downward, partly offsetting the optimistic bias of permissive defaults discussed below; we cannot quantify the net effect on RQ\textsubscript{3}, and the spot-check itself relies on a single annotator.

A related threat concerns the input given to scripts. Scripts receive the full before/after content of changed files, but not a line-level diff (Section~\ref{sec:rq3}). Instructions whose compliance criterion concerns only what a commit added or modified are therefore evaluated on the final state of each touched file, so pre-existing violations in a file the agent touched may be counted against the agent, and pre-existing compliant content may mask a violation the agent introduced elsewhere.

Another threat is that compliance is recorded as a binary signal, which cannot capture partial or graded adherence; we treat this as a deliberate scope decision tied to our artifact-based observability criterion rather than an oversight. 

One more threat concerns reliance on the LLM-generated \textit{candidate observable behavior} field as the functional specification for script generation; this is partially mitigated by its validation during RQ\textsubscript{1} (Section~\ref{sec:rq1}), though that validation covers only the sampled subset. 
 
Finally, LLM-generated scripts tend to be permissive by default, reporting compliance absent contrary evidence; we mitigate the resulting risk of vacuous compliance by deriving instruction--commit applicability directly from each instruction's own validated fixture file patterns (Section~\ref{sec:rq3}) rather than from coarser, hand-authored category-level rules, though this filter is only as complete as the fixture set it is derived from. It also matches files by extension or conventional basename, so a commit touching any file of the relevant type counts as applicable even when the touched file is unrelated to the instruction's target; a more precise, content-aware applicability check would reduce the remaining risk of vacuous compliance.

\textbf{External validity} concerns generalizability.
Our findings are scoped to three agents---Copilot, Claude Code, and OpenAI Codex---and may not generalize to other coding agents. The \textit{strict agent subset} used for our analyses, while improving attribution precision, is likely imbalanced across agents, since some expose clean bot identities while others rely on indirect signals such as branch naming; this could skew per-agent comparisons toward agents with more reliably attributable commits. 

Our corpus is further limited to public GitHub repositories containing an ACF, excluding enterprise or private agentic workflows that may use these artifacts differently. The pace of change in agent tooling and ACF conventions also means our results reflect a snapshot in time and may not hold for future agent versions. 

Finally, we rely on open-weight, locally executed models (Qwen 3.5 9B for annotation, Qwen2.5-Coder-14B-Instruct for script generation) rather than frontier proprietary models; extraction quality and operationalization outcomes could differ with a different underlying model.

\textbf{Conclusion validity} concerns the reliability of statistical inferences.
Applicability filtering (Section~\ref{sec:rq3}) reduces the effective sample available per instruction, and instructions with zero applicable commits are excluded from the statistical analysis entirely, which may underpower comparisons involving rarer instruction types. Aggregating to one compliance rate per instruction before cross-group comparison guards against pseudo-replication, since commits sharing a snapshot or an agent are not independent observations; however, this aggregation still yields coarse, high-variance rates for instructions evaluated against very few applicable commits. \revised{We assess this directly: restricting both cross-group comparisons to instructions with at least 3 and at least 5 applicable commits (Section~\ref{sec:rq3_sensitivity}) leaves the bimodal compliance split and the Copilot-versus-Claude~Code gap intact, but the \textit{Development Process} category deficit and the Copilot-versus-OpenAI~Codex gap do not survive Bonferroni correction at either threshold, indicating these two specific comparisons are underpowered rather than firmly established.}

Group sizes across categories and agents are also likely imbalanced, which we account for by using Kruskal-Wallis and post-hoc Mann-Whitney U tests with Bonferroni correction and Cliff's delta effect sizes (Section~\ref{sec:rq3}), rather than assuming normality or equal-sized groups. 

Finally, the RQ\textsubscript{1} dataset underlying all downstream statistics carries residual labeling noise: validation was performed by a single annotator, following \citet{ortloff2023different}'s rationale for well-bounded coding tasks, and cross-model agreement between the two candidate LLM annotators reached only 76.1\% on the observable label; this measurement error propagates into the RQ\textsubscript{2} and RQ\textsubscript{3} results built on top of it.
\section{Conclusion}
This paper investigated whether instructions documented in agent context files (ACFs) are reflected in the behavior of autonomous coding agents, structured around three research questions covering instruction characterization, operationalization, and compliance. Analyzing 621 ACF snapshots and 6{,}541 agent-generated commits across 268 repositories, we found that 6{,}552 instructions are explicit, actionable, and observable; that 62.9\% of these can be operationalized into executable measurement scripts; and that, among the 2{,}053 instructions ultimately evaluable against real commits, mean compliance is 59.8\%, with a strongly bimodal split between instructions followed universally and instructions never followed. Compliance varies significantly by instruction category and by originating agent, indicating that adherence to ACF guidance is neither uniform nor automatic.

This study is scoped to publicly available repositories using three widely adopted agents and relies on LLM-assisted extraction and generation validated through manual sampling; results should be interpreted within this context rather than as universal claims about agent behavior.

Future work will extend this pipeline to additional agents and a broader set of ACF conventions, investigate the underlying causes of category- and agent-level compliance gaps, and explore mechanisms for enforcing ACF adherence at generation time rather than only measuring it after the fact.

\begin{comment}
\section*{Data Availability}
The data that support the findings of this study are openly available in our online appendix \cite{appendix}.
\end{comment}

\section*{Acknowledgment}
We gratefully acknowledge the financial support of JSPS KAKENHI (JP24K02921, JP25K03100), JST ASPIRE (JPMJAP2415), and JST CREST (JPMJCR23M1, JPMJCR26X7). We also acknowledge the use of GPT and Claude to ensure the manuscript's linguistic accuracy and readability and to partially generate and correct the data collection and analysis scripts.
%%
%% The next two lines define the bibliography style to be used, and
%% the bibliography file.
\bibliographystyle{ACM-Reference-Format}
\bibliography{references}

@article{chatlatanagulchai2025agent,
  title={Agent READMEs: An empirical study of context files for agentic coding},
  author={Chatlatanagulchai, Worawalan and Li, Hao and Kashiwa, Yutaro and Reid, Brittany and Thonglek, Kundjanasith and Leelaprute, Pattara and Rungsawang, Arnon and Manaskasemsak, Bundit and Adams, Bram and Hassan, Ahmed E and others},
  journal={ACM Transactions on Software Engineering and Methodology},
  year={2025},
  publisher={ACM New York, NY}
}

@misc{li2025aiteammates,
      title={The Rise of AI Teammates in Software Engineering (SE) 3.0: How Autonomous Coding Agents Are Reshaping Software Engineering}, 
      author={Hao Li and Haoxiang Zhang and Ahmed E. Hassan},
      year={2025},
      eprint={2507.15003},
      archivePrefix={arXiv},
      primaryClass={cs.SE},
      url={https://arxiv.org/abs/2507.15003}, 
}

@article{ouyang2025empirical,
  title={An empirical study of the non-determinism of chatgpt in code generation},
  author={Ouyang, Shuyin and Zhang, Jie M and Harman, Mark and Wang, Meng},
  journal={ACM Transactions on Software Engineering and Methodology},
  volume={34},
  number={2},
  pages={1--28},
  year={2025},
  publisher={ACM New York, NY}
}

@incollection{treude2024qualitative,
  title={Qualitative data analysis in software engineering: Techniques and teaching insights},
  author={Treude, Christoph},
  booktitle={Handbook on Teaching Empirical Software Engineering},
  pages={155--176},
  year={2024},
  publisher={Springer}
}

@inproceedings{sasaki2024systematic,
  title={Systematic literature review of prompt engineering patterns in software engineering},
  author={Sasaki, Yuya and Washizaki, Hironori and Li, Jialong and Sander, Dominik and Yoshioka, Nobukazu and Fukazawa, Yoshiaki},
  booktitle={2024 IEEE 48th Annual Computers, Software, and Applications Conference (COMPSAC)},
  pages={670--675},
  year={2024},
  organization={IEEE}
}

@inproceedings{della2026toward,
  title={Toward measuring prompt quality: A preliminary investigation on prompt smells},
  author={Della Porta, Antonio and Voria, Gianmario and Abbate, Andrea and Sulipano, Raffaele and Lambiase, Stefano and Catolino, Gemma and Palomba, Fabio},
  booktitle={2026 IEEE International Conference on Software Analysis, Evolution and Reengineering-Companion (SANER-C)},
  pages={293--300},
  year={2026},
  organization={IEEE}
}

@inproceedings{ahmed2025can,
  title={Can LLMs replace manual annotation of software engineering artifacts?},
  author={Ahmed, Toufique and Devanbu, Premkumar and Treude, Christoph and Pradel, Michael},
  booktitle={2025 IEEE/ACM 22nd International Conference on Mining Software Repositories (MSR)},
  pages={526--538},
  year={2025},
  organization={IEEE}
}

@inproceedings{tan2024large,
  title={Large language models for data annotation and synthesis: A survey},
  author={Tan, Zhen and Li, Dawei and Wang, Song and Beigi, Alimohammad and Jiang, Bohan and Bhattacharjee, Amrita and Karami, Mansooreh and Li, Jundong and Cheng, Lu and Liu, Huan},
  booktitle={Proceedings of the 2024 conference on empirical methods in natural language processing},
  pages={930--957},
  year={2024}
}

@article{huang2025professional,
  title={Professional Software Developers Don't Vibe, They Control: AI Agent Use for Coding in 2025},
  author={Huang, Ruanqianqian and Reyna, Avery and Lerner, Sorin and Xia, Haijun and Hempel, Brian},
  journal={arXiv preprint arXiv:2512.14012},
  year={2025}
}

@article{watanabe2025use,
  title={On the use of agentic coding: An empirical study of pull requests on github},
  author={Watanabe, Miku and Li, Hao and Kashiwa, Yutaro and Reid, Brittany and Iida, Hajimu and Hassan, Ahmed E},
  journal={ACM Transactions on Software Engineering and Methodology},
  publisher={ACM New York, NY}
}

@article{wang2025agents,
  title={Agents in software engineering: Survey, landscape, and vision},
  author={Wang, Yanlin and Zhong, Wanjun and Huang, Yanxian and Shi, Ensheng and Yang, Min and Chen, Jiachi and Li, Hui and Ma, Yuchi and Wang, Qianxiang and Zheng, Zibin},
  journal={Automated Software Engineering},
  volume={32},
  number={2},
  pages={70},
  year={2025},
  publisher={Springer}
}

@article{hassan2025agentic,
  title={Agentic software engineering: Foundational pillars and a research roadmap},
  author={Hassan, Ahmed E and Li, Hao and Lin, Dayi and Adams, Bram and Chen, Tse-Hsun and Kashiwa, Yutaro and Qiu, Dong},
  journal={arXiv preprint arXiv:2509.06216},
  year={2025}
}

@inproceedings{bouzenia2025repairagent,
  title={Repairagent: An autonomous, llm-based agent for program repair},
  author={Bouzenia, Islem and Devanbu, Premkumar and Pradel, Michael},
  booktitle={2025 IEEE/ACM 47th International Conference on Software Engineering (ICSE)},
  pages={2188--2200},
  year={2025},
  organization={IEEE}
}

@inproceedings{feldt2023towards,
  title={Towards autonomous testing agents via conversational large language models},
  author={Feldt, Robert and Kang, Sungmin and Yoon, Juyeon and Yoo, Shin},
  booktitle={2023 38th IEEE/ACM International Conference on Automated Software Engineering (ASE)},
  pages={1688--1693},
  year={2023},
  organization={IEEE}
}

@article{horikawa2025agentic,
  title={Agentic Refactoring: An Empirical Study of AI Coding Agents},
  author={Horikawa, Kosei and Li, Hao and Kashiwa, Yutaro and Adams, Bram and Iida, Hajimu and Hassan, Ahmed E},
  journal={arXiv preprint arXiv:2511.04824},
  year={2025}
}

@article{shao2024llms,
  title={Are llms correctly integrated into software systems?},
  author={Shao, Yuchen and Huang, Yuheng and Shen, Jiawei and Ma, Lei and Su, Ting and Wan, Chengcheng},
  journal={arXiv preprint arXiv:2407.05138},
  year={2024}
}

@inproceedings{shrivastava2023repository,
  title={Repository-level prompt generation for large language models of code},
  author={Shrivastava, Disha and Larochelle, Hugo and Tarlow, Daniel},
  booktitle={International Conference on Machine Learning},
  pages={31693--31715},
  year={2023},
  organization={PMLR}
}

@inproceedings{le2025impacts,
  title={On the impacts of contexts on repository-level code generation},
  author={Le Hai, Nam and Nguyen, Dung Manh and Bui, Nghi DQ},
  booktitle={Findings of the Association for Computational Linguistics: NAACL 2025},
  pages={1496--1524},
  year={2025}
}

@inproceedings{chatlatanagulchai2025use,
  title={On the use of agentic coding manifests: An empirical study of claude code},
  author={Chatlatanagulchai, Worawalan and Thonglek, Kundjanasith and Reid, Brittany and Kashiwa, Yutaro and Leelaprute, Pattara and Rungsawang, Arnon and Manaskasemsak, Bundit and Iida, Hajimu},
  booktitle={Proceedings of the 27th International Conference on Product-Focused Software Process Improvement (PROFES)},
  pages={543--551},
  year={2025},
  organization={Springer}
}

@article{mohsenimofidi2025context,
  title={Context engineering for AI agents in open-source software},
  author={Mohsenimofidi, Seyedmoein and Galster, Matthias and Treude, Christoph and Baltes, Sebastian},
  journal={arXiv preprint arXiv:2510.21413},
  year={2025}
}

@article{sabra2025assessing,
  title={Assessing the Quality and Security of AI-Generated Code: A Quantitative Analysis},
  author={Sabra, Abbas and Schmitt, Olivier and Tyler, Joseph},
  journal={arXiv preprint arXiv:2508.14727},
  year={2025}
}

@article{hou2024large,
  title={Large language models for software engineering: A systematic literature review},
  author={Hou, Xinyi and Zhao, Yanjie and Liu, Yue and Yang, Zhou and Wang, Kailong and Li, Li and Luo, Xiapu and Lo, David and Grundy, John and Wang, Haoyu},
  journal={ACM Transactions on Software Engineering and Methodology},
  volume={33},
  number={8},
  pages={1--79},
  year={2024},
  publisher={ACM New York, NY}
}

@inproceedings{ortloff2023different,
  title={Different researchers, different results? analyzing the influence of researcher experience and data type during qualitative analysis of an interview and survey study on security advice},
  author={Ortloff, Anna-Marie and Fassl, Matthias and Ponticello, Alexander and Martius, Florin and Mertens, Anne and Krombholz, Katharina and Smith, Matthew},
  booktitle={Proceedings of the 2023 CHI Conference on Human Factors in Computing Systems},
  pages={1--21},
  year={2023}
}

@inproceedings{mathews2024test,
  title={Test-driven development and llm-based code generation},
  author={Mathews, Noble Saji and Nagappan, Meiyappan},
  booktitle={Proceedings of the 39th IEEE/ACM International Conference on Automated Software Engineering},
  pages={1583--1594},
  year={2024}
}

@article{fakhoury2024llm,
  title={Llm-based test-driven interactive code generation: User study and empirical evaluation},
  author={Fakhoury, Sarah and Naik, Aaditya and Sakkas, Georgios and Chakraborty, Saikat and Lahiri, Shuvendu K},
  journal={IEEE Transactions on Software Engineering},
  volume={50},
  number={9},
  pages={2254--2268},
  year={2024},
  publisher={IEEE}
}

@misc{chen2022codetcodegenerationgenerated,
      title={CodeT: Code Generation with Generated Tests}, 
      author={Bei Chen and Fengji Zhang and Anh Nguyen and Daoguang Zan and Zeqi Lin and Jian-Guang Lou and Weizhu Chen},
      year={2022},
      eprint={2207.10397},
      archivePrefix={arXiv},
      primaryClass={cs.CL},
      url={https://arxiv.org/abs/2207.10397}, 
}

@article{jiang2026survey,
  title={A survey on large language models for code generation},
  author={Jiang, Juyong and Wang, Fan and Shen, Jiasi and Kim, Sungju and Kim, Sunghun},
  journal={ACM Transactions on Software Engineering and Methodology},
  volume={35},
  number={2},
  pages={1--72},
  year={2026},
  publisher={ACM New York, NY}
}

@misc{galster2026configuringagenticaicoding,
      title={Configuring Agentic AI Coding Tools: An Exploratory Study}, 
      author={Matthias Galster and Seyedmoein Mohsenimofidi and Jai Lal Lulla and Muhammad Auwal Abubakar and Christoph Treude and Sebastian Baltes},
      year={2026},
      journal={3rd ACM/IEEE International Conference on AI-powered Software (AIware 2026)},
      eprint={2602.14690},
      archivePrefix={arXiv},
      primaryClass={cs.SE},
      url={https://arxiv.org/abs/2602.14690}, 
}

@misc{lulla2026impactagentsmdfilesefficiency,
      title={On the Impact of AGENTS.md Files on the Efficiency of AI Coding Agents}, 
      author={Jai Lal Lulla and Seyedmoein Mohsenimofidi and Matthias Galster and Jie M. Zhang and Sebastian Baltes and Christoph Treude},
      year={2026},
      eprint={2601.20404},
      archivePrefix={arXiv},
      primaryClass={cs.SE},
      url={https://arxiv.org/abs/2601.20404}, 
}

@misc{wang2025codeifbench,
      title={CodeIF-Bench: Evaluating Instruction-Following Capabilities of Large Language Models in Interactive Code Generation},
      author={Peiding Wang and Li Zhang and Fang Liu and Lin Shi and Minxiao Li and Bo Shen and An Fu},
      year={2025},
      eprint={2503.22688},
      archivePrefix={arXiv},
      primaryClass={cs.SE},
      url={https://arxiv.org/abs/2503.22688},
}

@misc{qi2025agentif,
  title={Agentif: Benchmarking instruction following of large language models in agentic scenarios},
  author={Qi, Yunjia and Peng, Hao and Wang, Xiaozhi and Xin, Amy and Liu, Youfeng and Xu, Bin and Hou, Lei and Li, Juanzi},
  journal={arXiv preprint arXiv:2505.16944},
  year={2025}
}

@misc{ehsani2026where,
  title={Where do ai coding agents fail? an empirical study of failed agentic pull requests in github},
  author={Ehsani, Ramtin and Pathak, Sakshi and Rawal, Shriya and Al Mujahid, Abdullah and Imran, Mia Mohammad and Chatterjee, Preetha},
  booktitle={Proceedings of the 23rd International Conference on Mining Software Repositories},
  pages={807--811},
  year={2026}
}

@misc{yang2026firstlook,
  title={A First Look at Coding Agents' Compliance with AI Contribution Rules in Open-Source Communities},
  author={Yang, Wenhao and He, Runzhi and Zhou, Minghui},
  journal={arXiv preprint arXiv:2607.26819},
  year={2026}
}

@article{gloaguen2026evaluating,
  title={Evaluating AGENTS. md: Are repository-level context files helpful for coding agents?},
  author={Gloaguen, Thibaud and M{\"u}ndler-Sasahara, Niels and M{\"u}ller, Mark Niklas and Raychev, Veselin and Vechev, Martin},
  journal={arXiv preprint arXiv:2602.11988},
  year={2026}
}

@inproceedings{hora2026coding,
  title={Are coding agents generating over-mocked tests? an empirical study},
  author={Hora, Andre and Robbes, Romain},
  booktitle={Proceedings of the 23rd International Conference on Mining Software Repositories},
  pages={335--346},
  year={2026}
}

@inproceedings{ardic2026uncovering,
  title={Uncovering AGENTS. md: What Testing Guidance Open Source Projects Provide to Coding Agents},
  author={Ardic, Baris and Olsthoorn, Mitchell and Zaidman, Andy},
  booktitle={International Working Conference on Source Code Analysis and Manipulation (SCAM). IEEE, Benevento, Italy},
  year={2026}
}

@misc{replication_package,
    title={Replication Package},
    author={Voria, Gianmario and Kashiwa, Yutaro and Catolino, Gemma and Palomba, Fabio},
    url={https://github.com/gianwario/Agents-Context-Files-Analysis}
}

%%
%% If your work has an appendix, this is the place to put it.
%\appendix

\end{document}